\documentclass[article]{agujournal2019_arxiv}
\usepackage{url} 
\usepackage{lineno}
\usepackage{soul}

\usepackage{graphicx}
\usepackage{amsmath}
\usepackage{mathtools}

\usepackage{natbib}
\newcommand{\vect}[1]{\boldsymbol{#1}}
\usepackage{mathtools,bbm}

\usepackage{amssymb}
\usepackage{gensymb}

\usepackage[nomarkers,figuresonly]{endfloat}

\usepackage{amsmath}

\usepackage{bbm}

\usepackage{float,graphicx}
\usepackage{tabularx, booktabs}

\usepackage{xcolor}

\usepackage{algorithm}
\usepackage[noend]{algpseudocode}

\usepackage{footnote}

\algrenewcommand\algorithmiccomment[1]{\hfill \(\phantom{ }\) #1}

\newcommand{\algorithmfootnote}[2][\footnotesize]{%
  \let\old@algocf@finish\@algocf@finish
  \def\@algocf@finish{\old@algocf@finish
    \leavevmode\rlap{\begin{minipage}{\linewidth}
    #1#2
    \end{minipage}}%
  }%
}

\draftfalse

\journalname{BSSA}

\usepackage{hyperref}

\hypersetup{
    colorlinks,
    citecolor=black,
    linkcolor=black
}

\begin{document}

\title{Earthquake Phase Association \\ with Graph Neural Networks}

\authors{I.W. McBrearty, G.C. Beroza}

\affiliation{1}{Department of Geophysics, 397 Panama Mall, Stanford University, Stanford, California 94305-2215, USA} \vspace{0.2 cm}

\correspondingauthor{I.W. McBrearty}{imcbrear@stanford.edu}

\begin{keypoints}
\item Microseismicity Detection
\item Northern California Seismicity
\item \textbf{G}raph \textbf{E}arthquake \textbf{N}eural \textbf{I}nterpretation \textbf{E}ngine
\end{keypoints}


\begin{abstract}

Seismic phase association connects earthquake arrival time measurements to their causative sources.  Effective association must determine the number of discrete events, their location and origin times, and it must differentiate real arrivals from measurement artifacts. The advent of deep learning pickers, which provide high rates of picks from closely overlapping small magnitude earthquakes, motivates revisiting the phase association problem and approaching it using the methods of deep learning. We have developed a Graph Neural Network associator that simultaneously predicts both source space-time localization, and discrete source-arrival association likelihoods. The method is applicable to arbitrary geometry, time-varying seismic networks of hundreds of stations, and is robust to high rates of sources and input picks with variable noise and quality. Our Graph Earthquake Neural Interpretation Engine (GENIE) uses one graph to represent the station set and another to represent the spatial source region.  GENIE learns relationships from data in this combined representation that enable it to determine robust source and source-arrival associations. We train on synthetic data, and test our method on real data from the Northern California (NC) seismic network using input generated by the PhaseNet deep learning phase picker. We successfully re-detect $\sim$96$\%$ of all events M$>$1 reported by the USGS during 500 random days between 2000$-$2022. Over a 100-day continuous interval of processing in 2017$-$2018, we detect $\sim$4.2$\times$ the number of events reported by the USGS. Our new events have small magnitude estimates below the magnitude of completeness of the USGS catalog, and are located close to the active faults and quarries in the region. Our results demonstrate that GENIE can effectively solve the association problem under complex seismic monitoring conditions.

\end{abstract}


\section{Introduction}

Improved earthquake catalogs are essential for many purposes, including: understanding active fault structures, illuminating foreshock and aftershock behaviors, and developing better tomographic images of the subsurface. Building catalogs usually involves picking impulsive arrivals on individual seismic stations, \textit{associating} those arrivals to common sources (and assigning phase types to each arrival), followed by characterizing the individual earthquakes (e.g., location, origin time, magnitude, and moment tensor). The rapidly growing application of machine learning to earthquake detection and phase picking \citep{kong2019machine, bergen2019machine, mousavi2022deep} has led to greatly increased numbers of phase picks, producing results that differ from those measured by traditional methods in several important ways (e.g., \cite{meier2019reliable, zhu2019phasenet, mousavi2020earthquake, xiao2021siamese}).  Unlike traditional methods, machine learning methods often provide equal numbers of P and S picks, and in many cases can classify the wave type (P or S) of phase measurements. Machine learning based pickers also often pick many more arrivals, such that the arrival times of multiple sources frequently overlap in time (e.g., Fig. 1). This time entanglement of information from multiple sources can lead to ambiguities that make phase association challenging. Recent dense earthquake catalogs in southern California \citep{ross2019searching}, northern Chile \citep{mcbrearty2019earthquake}, Central US \citep{park2020machine, park2022basement}, Italy \citep{tan2021machine}, and Indonesia \citep{jiang2022detailed} indicate there may be thousands of detectable wave arrivals per day from seismic events in tectonically active regions. These high rates of picks are largely from earthquakes below the magnitude of completeness of traditional catalogs. The ability to use these measurements to generate improved, reliable catalogs, depends critically on the ability to solve the phase association problem accurately.

Associators are typically based on the back-projection paradigm (e.g., \citet{johnson1997robust, ringdal1989multi, draelos2015new}) to localize sources through direct time reversal and stacking. Recent studies have included: deep-learning-based RNN's to predict association likelihoods between arrivals without inverting for a source \citep{ross2019phaselink}, probabilistic \citep{le2020net} and maximum-likelihood models \citep{sheen2021seismic}, refined back-projection and post-processing schemes \citep{yeck2019glass3, zhang2019rapid, chowdhury2021faser}, back-projection and graph-based optimization \citep{mcbrearty2019earthquake}, random sample and consensus clustering \citep{woollam2020hex, zhu2021multi}, Bayesian mixture model \citep{zhu2022earthquake}, and pairwise feature-based \citep{bergen2018detecting} and deep-learning-based \citep{mcbrearty2019pairwise, dickey2020beyond} waveform associations. A growing set of models are emerging to circumvent the association problem, by training models to map raw waveforms directly to source predictions in the so-called end-to-end fashion \citep{van2020automated, zhang2020locating, munchmeyer2021earthquake, shen2021array, yang2021simultaneous, zhu2022end}. While these approaches have highlighted the significant potential for improving phase association, two notable findings stand out. The first is that deep learning \citep{ross2019phaselink} offers a powerful framework to formalize the prediction problems of interest, and allows learning mappings from data that are potentially more expressive than those based on hand-crafted features. The second is that graphs \citep{mcbrearty2019earthquake, yeck2019glass3} offer a convenient means for organizing the relevant physical elements of \textit{\{picks, stations, source region\}} into a data structure that elucidates the essential interactions.

In this study we present a solution to the association problem that combines the strengths of deep learning with the strengths of graphs by using Graph Neural Networks (GNNs). By using physically inspired adjacency matrices at numerous layers inside the GNN, our method directly exploits key information, such as station proximity and travel time moveout across a seismic network, while processing the input data and making predictions of the continuous source space-time likelihoods, and discrete source-arrival association likelihoods. The resulting model effectively associates phases even with high rates of events, high rates of false picks, overlapping arrival times, and can handle large aperture seismic networks. It is trained to allow the seismic network to change in time (the number of stations, and positions, can vary), can handle hundreds of stations simultaneously, and allows each station to have any number of picks per time window. The rates of false picks, station coverage, and quality of sensors can also be highly variable across the network. The architecture we propose features distinct sub-modules, each of which uses different adjacency matrices (or equivalently, edge lists) that are designed to represent and capture information about the essential components of the problem: the station network, spatial domain, theoretical travel times, pick dataset, and prediction tasks (source location, timing, and pairwise source-arrival association likelihoods). Fig. 2 highlights many of these concepts.

Several of the design choices we make, such as using pairwise information between nearby stations (for a fixed source), and pairs of nearby sources (for a fixed station), are reminiscent of traditional processing methods such as geometric reverse-time migration \citep{nakata2016reverse}, sub-network processing approaches \citep{wu2002virtual}, and double-difference earthquake relocation \citep{waldhauser2000double}. The difference from these traditional processing methods is that rather than directly implement these concepts, they are implicitly built into the space of learnable functions over which the GNN is trained. The edges of graphs we define connect relevant physically entities, and we learn graph convolution operators that perform well over these structures for the prediction tasks at hand. This highlights the potential of GNN's to easily allow domain experts to inject intuition into their neural networks, guiding the inductive bias \citep{battaglia2018relational} to focus on the most important functional inter-dependencies.

\subsection{The Association Challenge}

Seismic phase association is the problem of determining which arrivals picked across a seismic network are from a common source, correctly grouping these picks for each source, and assigning phase types to each arrival (Figs. 1,2). Automating this procedure for general seismic networks has remained a subtle and persistent challenge for decades (e.g., \cite{ringdal1989multi, johnson1997robust}). When applied to large aperture networks with high rates of seismicity, the essential challenges include:

\indent

\begin{itemize}
\item{\textbf{The number of earthquakes (and their locations) are unknown}}
\item{\textbf{Many picks can be false, with uncertain pick times}}
\item{\textbf{Multiple earthquakes can occur nearby in time and space}}
\item{\textbf{Dozens or hundreds of stations must be processed}}
\item{\textbf{Station coverage, and source distribution, are highly heterogeneous}}
\item{\textbf{Most events are small and only observed on a small subset of stations}}
\item{\textbf{The available seismic network changes in time}}
\end{itemize}

Any comprehensive, successful association algorithm needs to address all of these complications. Recently proposed associators have made great strides in handling some of these challenges; however, high event rates, and overlapping events remain a challenge. Few methods have also directly addressed the major issue of large aperture networks, which is that each small earthquake will produce arrivals on only the nearest handful of stations, while the full network may contain hundreds of stations with abundant picks from other sources at similar times. This is problematic for time reversal and back-projection-based methods because it frequently leads to spurious alignment of real earthquake moveouts with false picks that should be categorized as unassociated picks, yet the travel time information alone cannot make this discrimination. Many methods also suffer from sensitivity to temporal changes in seismic network geometry and station quality, which can obfuscate the true behavior of seismicity. Methods should hence strive to be insensitive to temporally changing seismic network coverage.

\subsection{The Case for Graph Neural Networks}

Graph neural networks (GNN's) are a class of deep learning (DL) models that effectively solve \textit{heterogenous} and \textit{non-Euclidean} data processing tasks (e.g., \citet{hamilton2017inductive, battaglia2018relational}). In the machine learning community, heterogeneous and non-Euclidean describe the wide class of real world problems where the observed data is more naturally represented as residing on an arbitrary manifold, or more simply, as a diverse set of measurements on an irregular collection of nodes (e.g., seismometers), for which there is no canonical grid ordering, or grid-topology to represent the data. Images and time series are examples of Euclidean data (residing on regular grids); however, collections of discrete picks on an irregular seismic network are non-Euclidean (there is no canonical ordering of arbitrarily positioned stations). Nearly all of the traditional DL-models (e.g., Fully Connected Networks (FCN's), Convolutional Neural Networks (CNN's), Recurrent Neural Networks (RNN's)), are designed to be effective for Euclidean data, and practitioners often make difficult and heuristic choices in how to implement FCNs, CNNs, RNNs, on inherently non-Euclidean tasks (Fig. 3). The association problem, which combines processing discrete picks across irregular seismic network geometries, and inferring the number (and location) of sources, combined with assigning all arrivals to their respective sources (and phase types), is a quintessential heterogeneous processing task.

GNN's excel at processing non-Euclidean and heterogeneous data because customized graphs can be introduced to represent non-Euclidean data structures \citep{zhou2020graph}, and the neural network operators are designed and trained specifically to use the underlying graph (or topology) to make predictions. This allows GNN's to use the inherent structure of the problem as the computational backbone, such that the processing occurs over the real non-Euclidean domain (or a close approximation to it), rather than artificially forcing non-Euclidean problems into a Euclidean framework. This can be thought of as a form of regularization, where the \textit{inductive bias} \citep{battaglia2018relational} that nearby nodes on the graph should interact more than distant nodes on the graph while making predictions is implicitly built into the architecture (through the adjacency matrix). From another perspective, making use of an adjacency matrix for data processing is similar to using a full covariance matrix (with off diagonal terms included) in Bayesian inverse theory: the adjacency matrix helps illuminate the different pairwise interactions between various correlated data contained in the full dataset, in much the same way that a covariance matrix can help account for correlations among data. This, and other connections between graphs and traditional fields in data science are clarified in the framework of graph signal processing \citep{djuric2018cooperative, ortega2018graph}, which is a generalization of standard signal processing tools such as convolution, filtering, decimation, shifting, and Fourier transform, onto irregular, non-Euclidean domains. Due to their many strengths, we anticipate that GNNs, and graph signal processing in general, will prove to be productive approaches across the geosciences.

\subsection{The Essentials: Input and Output}

Our graph neural network is designed to work on variable geometry seismic networks, where each input can accept a different station set $\mathcal{S} = \{\vect{s}_i \in {\Omega_{\mathcal{X}}} \mid i \leq N\}$, with station positions $\vect{s}_i$ within a spatial domain $\Omega_{\mathcal{X}}$. The input pick data is a variable discrete set of sets, $\mathcal{D} = \{\mathcal{D}_i \mid  i \leq |\mathcal{S}| \}$, where each station has pick set list $\mathcal{D}_i = \{\tau_1, \tau_2, \cdots \tau_{M_i}\}$, with a variable number of picks per station, $M_i$. Additionally, each pick can be tagged with phase type, amplitude, or azimuth information. Our initial application is focused on the simplest type of input, which is a set of unlabelled arrival times without supplementary information or phase type. We let $T_k(\vect{s}, \vect{x})$ represent theoretical travel times from source ($\vect{x}$) to station ($\vect{s}$), with phase types P ($k = 1)$ and S ($k = 2)$ waves; however, the method treats each phase type in a simple parallel scheme that can be applied to any number of phase types.

For the given input, $\mathcal{D} = \{\mathcal{D}_i\}$, with arbitrary $\mathcal{S}$, the GNN produces a continuous, bounded, and differentiable spatio-temporal prediction of source likelihood throughout $\Omega = \Omega_{\mathcal{X}} \times [0, W]$, where $W = 10$ s in our applications. It also provides a prediction of any pairwise, source-arrival (and phase dependent) likelihood of association between source queries $(\vect{x}, t) \in \Omega$ and picks across any station $\tau \in \mathcal{D}$. What this means is that for any source ($\vect{x}, t)$, with $t \in [0, W]$, we can query the association likelihood of all picks in the window $[0, \text{max}_{\vect{s}}$ $ T(\vect{s}, \vect{x}) + W]$ to this source, and the predictions change smoothly as a function of both source coordinates and origin time (throughout the spatial region $\Omega_{\mathcal{X}}$, and over the time interval $W$). To process over continuous time, we step forward by a fraction of $W$, and stack overlapping portions of the predictions. In the flow of the neural network, the source-arrival association predictions are made after the GNN has made spatio-temporal source predictions internally (by predicting $p(\vect{x},t) \subset \Omega_{\mathcal{X}} \times [0,W]$), allowing the association predictions to be conditioned by the predictions already obtained over the spatio-temporal domain. In other words, the GNN first decides the sources that are active, then assigns likelihoods of pick associations to these sources (Fig. 2), and these mappings occur in a continuous way, sequentially after one another.

Hence, the network achieves the triple tasks of

\begin{subequations} \label{eq:functions}
\begin{equation}
    f_1 : \bigl(\mathcal{D}_{i \leq |\mathcal{S}|}, \mathcal{S} \bigr) \longrightarrow \bigl\{[0,1]^{|\mathcal{X}| \times [0,W]}, f_2, f_3\bigr\}
\end{equation}
\begin{equation}
    f_2 : \bigl((\vect{x},t) \in \Omega) \longrightarrow \bigl[0,1\bigr]
\end{equation}
\begin{equation}
    f_3 : \bigl((\vect{x},t) \in \Omega, \tau \in \mathcal{D}\bigr) \longrightarrow \bigl[0,1\bigr]
\end{equation}
\end{subequations}

\noindent in which $f_1$ maps the arbitrary pick datasets, and arbitrary seismic network into continuous-time predictions over the discrete spatial grid $\mathcal{X} \times [0,W]$, while also producing two new functions, $f_2$, and $f_3$, where $f_2$ predicts spatio-temporal predictions of source likelihood at any continuous spatio-temporal coordinate $(\vect{x},t) \in \Omega$, and $f_3$ predicts association likelihood for any pair of sources $(\vect{x},t) \in \Omega$, and arrivals $\tau \in \mathcal{D}$. In practice, this partitioning of the forward map \eqref{eq:functions} allows the network to be used in targeted ways: the first produces a reference prediction over the static grid $\mathcal{X}$, and the second, $f_2$, can be used to refine initial source predictions in continuous $\Omega_{\mathcal{X}}$ with a dense grid search, multi-step hierarchical search, or continuous gradient descent (using automatic differentiation of $f_2$ with respect to the input). Finally, once optimal sources are obtained, $f_3$ is queried to determine which arrivals $\tau \in \mathcal{D}$ have high association likelihood to these sources (or conversely, to establish which arrivals have low association likelihood to queried sources). Through these steps the GNN becomes an easy to use, general purpose associator. It flexibly allows changing seismic networks and dense pick data sets, while providing robust (and smooth), continuous predictions of source and source-arrival association likelihoods. 

\section{Methods}

\subsection{Sampling Input with Backprojection Paradigm}

To extract an explicit input tensor from $\mathcal{D}$ that can be passed to the GNN, we use the simple physically inspired heuristic of back-projection \citep{ringdal1989multi}. Since a source at $(\vect{x}, t_0)$ will theoretically produce arrivals at times $t_0 + T_k(\vect{s}, \vect{x})$, for all stations $\vect{s}_i \in \mathcal{S}$, the simplest input tensor to extract from $\mathcal{D}$ that can illuminate the presence of sources is to compute

\begin{equation} \label{eq: input tensor}
    \vect{h}_{k}^0(\vect{s}_i, \vect{x}) = \max\limits_{\tau_i \in \mathcal{D}_i} \Biggl[ \exp\Biggl(-\frac{(t_0 + T_k(\vect{s}_i, \vect{x}) - \tau_i)^2}{2\sigma^2}\Biggr) \Biggr]
\end{equation}

\noindent for a given query time $t_0$, which is simply a Gaussian kernel applied to the nearest neighbor lookup of the closest arrival on each station $i$, to the theoretical arrival time of $t_0 + T_k(\vect{s}_i, \vect{x})$ of phase type $k$, for a given source $\vect{x}$. This is nearly equivalent to the traditional `pre-stack' BP metric \citep{ringdal1989multi}. In our scheme, we collect this measurement \eqref{eq: input tensor} for both phase types $k = 1$ (P waves) and $k = 2$ (S waves), on all nodes of a modest size (e.g., 250 nodes) set, $\vect{x} \in \mathcal{X}$, where $\mathcal{X}$ is any sparse, space filling, dense packing of $\Omega_{\mathcal{X}}$. This tensor $\vect{h}_k^0(\vect{s}, \vect{x}) \in \mathbb{R}^{|\mathcal{S} \times \mathcal{X}| \times 2}$ will have entries given by the Cartesian product of $\mathcal{S} \times \mathcal{X}$ (e.g., 200 stations $\times$ 250 source nodes = 50,000 entries). A key design choice in our GNN is to apply graph convolutions directly to this object, on the data measured over the pairwise set, $\mathcal{S} \times \mathcal{X}$, rather than first `summing' or `stacking' over stations, or using any other additional pre-processing. How we apply graph convolutions to $\mathcal{S} \times \mathcal{X}$ directly is explained throughout the next few sections.

More sophisticated input schemes than defined in \eqref{eq: input tensor} could include passing in a sum over nearby picks, or passing in the full set of picks for each station-source pair. These operations are defined as

\begin{subequations}
\begin{equation} \label{eq: input type 1}
    \vect{h}^{0}_k(\vect{s}_i, \vect{x}) = \sum\limits_{\tau_i \in \mathcal{D}_i} \Biggl[ \exp\Biggl(-\frac{(t_0 + T_k(\vect{s}_i, \vect{x}) - \tau_i)^2}{2\sigma^2}\Biggr) \Biggr]
\end{equation}
\begin{equation} \label{eq: input type 2}
    \vect{h}^{0}_k(\vect{s}_i, \vect{x}) = \Biggl\{ \exp\Biggl(-\frac{(t_0 + T_k(\vect{s}_i, \vect{x}) - \tau_i)^2}{2\sigma^2}\Biggr), \vect{e} \mid \tau_i \in \mathcal{D}_i \Biggr\}.
\end{equation}
\end{subequations}

\noindent These variants would have the advantage of passing more complete information about the raw pick data to the GNN, at the expense of increased IO and computing costs. Importantly, however, it is only the third option \eqref{eq: input type 2} where the full set of adjacent picks (to the theoretical time) is passed in as a set, with all relevant edge data ($\vect{e}$) included, which avoids the loss of information that otherwise occurs. The max-pool \eqref{eq: input tensor} and sum-pool \eqref{eq: input type 1} schemes both lead to loss of information immediately upon input, so a direction for improvement would be to implement variant three, despite the increased cost, which would allow the GNN greater access into the full pick dataset.

\subsection{Graph Message Passing}

The basic computational unit of most modern GNN's is the generic \textit{message passing} layer. The message passing layer encapsulates the powerful notion of \textit{graph convolution} (e.g., \citet{hamilton2017inductive,  battaglia2018relational}), wherein the latent states of nodes on graphs are updated based on learned mappings from the latent states of neighboring nodes. While various types of graph convolution have been designed over time, such as local-mean pooling \citep{kipf2016semi}, spectral methods \citep{defferrard2016convolutional}, and attention style mechanisms \citep{velivckovic2017graph}, it is arguably the concept of applying local pooling and a concatenation (or residual connection) between subsequent layers (e.g. \citet{hamilton2017inductive}) that have significantly improved the stability of training deep GNN's and increased widespread adoption. Other important advances have been increasing the total expressible information content of node embeddings (e.g., \citet{you2019position, you2021identity}), refined training practices \citep{zhou2020effective}, and designing effective software \citep{fey2019fast}.

The general form of a graph convolution layer is given by

\begin{equation} \label{eq:convolution cell}
    \vect{h}^{(k + 1)}_i = \phi^{agg}(\vect{h}^{(k)}_i, \text{POOL} \{\phi^{msg}(\vect{h}^{(k)}_j, \vect{e}_{ij}, \vect{z}) \mid j \in \mathcal{N}(i)\})
\end{equation}

\noindent where $\mathcal{N}(i)$ is the neighborhood of node $i$, $\vect{h}^{(k)}_i$ is the feature vector on node $i$, $\vect{e}_{ij}$ is edge data (e.g., cartesian offset) between nodes $i,j$, and $\phi^{msg}$ and $\phi^{agg}$ are distinct fully connected neural network layers. This operation is simply a parallel application of the same FCN neural network, $\phi^{msg}$, applied to all nodes of each neighborhood set, $\mathcal{N}(i)$, aggregation of the transformed messages in each neighborhood via \textit{pooling} (i.e., mean, max, or sum-pool) over each node's local neighborhood, and an additional FCN transformation, $\phi^{agg}$, to the returned aggregate message and the current node state, $\vect{h}_i^{(k)}$. An optional global term, $\vect{z} = \text{POOL}_i\{\phi^{glb}(\vect{h}_i)\}$, can be included to augment the message and aggregation functions with additional long range summary information over the entire graph signal. In \eqref{eq:convolution cell}, the `comma operator' represents concatenation across the feature dimension. This compact representation of the forward map encourages the learned FCN's to use both regular feature transformations and the adjacency matrix structure to guide the predictions.

\subsection{Spatial Adjacency Matrices}

Our graph neural network architecture is designed to use one component that specifies the station set, $\mathcal{S}$ and another that specifies the source region, $\mathcal{X}$. In the general case, the station set $\mathcal{S}$ can be an arbitrary realization, with any number of stations with arbitrary positions. Similarly, $\mathcal{X}$ can be a sparse grid of spatial points, that roughly fill the space of $\Omega_{\mathcal{X}}$ (i.e., a dense packing of 100's of spatial nodes, obtainable with K-means clustering). Each of the points $\vect{x} \in \mathcal{X}$ represent one spatial coordinate where we sample from the input data $\mathcal{D}$ at the lag times given by the moveout vector of that source, $t_0 + T_k(\vect{s}, \vect{x})$, in the manner specified in \eqref{eq: input tensor}. Rather than leave $\mathcal{S}$ and $\mathcal{X}$ as arbitrary unordered sets, it is essential to give these sets structure. This is done by interpreting these data as graphs: specifically, $\mathcal{G}_{\mathcal{S}} = (\mathcal{S}, \mathcal{E}_{\mathcal{S}})$ is the station graph, with node set $\mathcal{S}$ and edge list $\mathcal{E}_{\mathcal{S}}$, and $\mathcal{G}_{\mathcal{X}} = (\mathcal{X}, \mathcal{E}_{\mathcal{X}})$ is the spatial graph with node set $\mathcal{X}$ and edge list $\mathcal{E}_{\mathcal{X}}$. Unless the context requires additional precision, for notation purposes, we often refer to $\mathcal{S}$ and $\mathcal{X}$ instead of $\mathcal{G}_{\mathcal{S}}$ or $\mathcal{G}_{\mathcal{X}}$, where it is implied that $\mathcal{S}$ and $\mathcal{X}$ represent graphs and the edge structures $\mathcal{E}_{\mathcal{S}}$, $\mathcal{E}_{\mathcal{X}}$ are also available.

To complete the construction of station and spatial graphs, we need a \textit{scheme} to determine which edges to include in $\mathcal{E}_{\mathcal{S}}$ and $\mathcal{E}_{\mathcal{X}}$. There is no one way, or even one optimal adjacency matrix (equivalently, edge list), that is the best choice for all tasks. The optimal choice is problem dependent, and entire lines of research in representation learning  have been devoted to this complex question \citep{hamilton2017representation}. A major strength of GNNs; however, is that they typically work well for a range of adjacency matrix choices \citep{zhou2020graph}, because the forward map \eqref{eq:convolution cell} is trained to work with the distribution of adjacency matrices it is trained over. Hence, it is often sufficient to rely on simple heuristic schemes for determining the edge structure. For example, for point cloud data with physical reference coordinates, such as we have with $\mathcal{S}$ and $\mathcal{X}$, a natural choice is K-nearest-neighbor graphs, or $\epsilon-$distance graphs \citep{berton2018comparison}. Both constructions prefer \textit{spatially local} edge connections, which promote distributed, local computing, between small subsets of neighboring nodes, rather than dense long-range interactions between all of the nodes simultaneously. As K- or $\epsilon$-increases, the aperture of each graph convolution increases, much as kernel size in a CNN determines the width of the receptive field of the convolution. Repeated applications of graph convolution expand the effective receptive field of each node to paths on the graph (Fig. 4), of increasing `1-hop' distance, for each additional convolution.

\subsubsection{Edge Sets}

For our application we use $K\text{-}NN$ graphs, as these are slightly more robust than $\epsilon$-distance graphs to heterogeneously distributed point clouds, such as occurs with a typical seismic network station set, $\mathcal{S}$. They are also well behaved in that they have predictable levels of sparsity, as they fix the (incoming) degree of all nodes to K, whereas $\epsilon$-distance graphs can have substantially different degrees across different nodes of the graph. We note that $K\text{-}NN$ graphs of this kind are \textit{directed} (i.e., usually an edge $i \rightarrow j \in \mathcal{E}$ does not necessarily imply $j \rightarrow i \in \mathcal{E}$), and further these graphs can have some undesirable properties, for example that individual nodes can have unusually high outgoing edge degree (i.e., may be a nearby neighbor to many other nodes), and there are no guarantees on the global connectivity of the graph. While we have not optimized these choices, we obtained acceptable accuracy with the simple choice of $10$-nearest-neighbors and $15$-nearest-neighbors for $\mathcal{S}$ and $\mathcal{X}$, respectively. The edge lists of either graph are hence given for any input $\mathcal{S}$ and $\mathcal{X}$ as

\begin{subequations} \label{eq:edge connections}
\begin{equation}
    \mathcal{E}_{\mathcal{S}} = \{(i,j) \mid \vect{s}_j \in K\text{-}NN_{10}(\vect{s}_i)\}
\end{equation}
\begin{equation}
    \mathcal{E}_{\mathcal{X}} = \{(i,j) \mid \vect{x}_j \in K\text{-}NN_{15}(\vect{x}_i)\},
\end{equation}
\end{subequations}

\noindent where $K\text{-}NN_K$ are the K-nearest-neighbors for each input node. Examples of these graphs and the resulting edge connections are given in Fig. 1c,d. These nearest-neighbor lookup sets can be computed rapidly with GPU acceleration using the PyTorch Geometric package \citep{fey2019fast}, and hence allow computing these edge sets dynamically at the time the inputs are passed to the GNN and the forward call is implemented (as well as passing along any meta-data per edge, such as the relative locations of nodes).

We have specified the structure of $\mathcal{S}$ and $\mathcal{X}$, but we are still missing one essential graph. The raw data as measured by the BP metric \eqref{eq: input tensor} is first observed on the \textit{Cartesian product}, $\mathcal{S} \times \mathcal{X}$, of all pairs of sources and stations. Conveniently, the other two graphs, $\mathcal{G}_{\mathcal{S}}$ and $\mathcal{G}_{\mathcal{X}}$, can be used directly to define the formal Cartesian product graph, $\mathcal{G}_{\mathcal{S} \times \mathcal{X}} = (\mathcal{S} \times \mathcal{X}, \mathcal{E}_{\mathcal{S} \times \mathcal{X}})$, as a graph that has a node set of all distinct pairs of sources-stations, with edge structure that directly inherits the edge structures given by $\mathcal{E}_{\mathcal{S}}$ and $\mathcal{E}_{\mathcal{X}}$. Specifically, each $i$th node in $\mathcal{S} \times \mathcal{X}$ is a source-station pair, $(\vect{s}, \vect{x})_i$, and two nodes $(i,j)$ in $\mathcal{S} \times \mathcal{X}$ are linked by a directed edge if either

\begin{subequations} \label{eq: cartesian product graph}
\begin{equation}
    \mathcal{E}_{\mathcal{X} \leftarrow \mathcal{X}, \mathcal{S}} = \bigl\{ (i,j) \mid \vect{x}_j \in \mathcal{N}(\vect{x}_i) \wedge (\vect{s}_j = \vect{s}_i) \bigr\}
\end{equation}
\begin{equation}
    \mathcal{E}_{\mathcal{S} \leftarrow \mathcal{S}, \mathcal{X}} = \bigl\{ (i,j) \mid \vect{s}_j \in \mathcal{N}(\vect{s}_i) \wedge (\vect{x}_j = \vect{x}_i) \bigr\}.
\end{equation}
\end{subequations}

\noindent This edge scheme \eqref{eq: cartesian product graph} connects each (source-station) pair to other (source-station) pairs, for either (i) two nearby sources, and a fixed station, or (ii) two nearby stations, and a fixed source. These alternative relationships offer complementary insight into the raw feature data $\vect{h}_k^0(\vect{s}, \vect{x})$ observed on $\mathcal{S} \times \mathcal{X}$. Intuitively, while implementing graph convolutions on $\mathcal{E}_{\mathcal{S} \times \mathcal{X}}$, the two edge types allows us to process using the semantic concepts of `nearby stations', and `nearby sources should interact', without stacking over either axis explicitly. The (incoming) degree of each node in $\mathcal{G}_{\mathcal{S} \times \mathcal{X}}$ is still limited by the choice of K in either component graph, $\mathcal{G}_{\mathcal{S}}$ or $\mathcal{G}_{\mathcal{X}}$, and hence while the node set of $\mathcal{S} \times \mathcal{X}$ is very large (e.g., 200 stations x 250 source nodes = 50,000 nodes), the edge structure of this graph $\mathcal{E}_{\mathcal{S} \times \mathcal{X}}$ is still highly sparse (e.g., 10 edges / 50,000 nodes $=$ 2$\times 10^{-4}$ edge density per node), making sparse implementations of graph convolutions \citep{fey2019fast} on this graph tractable.

\subsection{Sketch of the Architecture}

\textbf{G}raph \textbf{E}arthquake \textbf{N}eural \textbf{I}nterpretation \textbf{E}ngine (\textbf{GENIE})

The GNN we use requires the assembly (or composition) of several distinct GNN networks that we refer to as modules (or sub-modules). This decomposition of the full forward map into several distinct modules allows us to control precisely the relevant data that interact at each step, and to gradually transform the raw pick data, $\mathcal{D}$, and station set, $\mathcal{S}$, to the final predictions. In order, the modules used to infer source predictions are: [1] Data Aggregation, [2] Bipartite Read-In, [3] Spatial Aggregation, [4] Spatial Attention, and [5] Temporal Attention. The modules used to infer source-arrival association predictions share the common initial [1-5] modules, and are followed by [6] Bipartite Read-Out, [7] Data Aggregation, [8] Arrival Embedding, and [9] Source-Station-Arrival Attention. In the supplemental materials, we define exactly each of the details of the `message passing' convolution cell (4) that is required to implement modules [1 - 9]; this includes specifying the edge sets, type of edge data available, number of repeated convolution applications per module, the feature dimensions of the individual FCN's, and the non-linear activation functions used. A visual depiction of the architecture is given in Fig. S1.

Inside the forward pass of the GNN, the sources are first used to sample the station pick data $\mathcal{D}_i$ at the theoretical lag times of all of the sources \eqref{eq: input tensor}; this creates the essential, and ubiquitous, raw feature $\vect{h}^0(\vect{s}, \vect{x})$, recorded on the node set of the Cartesian product graph, $\mathcal{S} \times \mathcal{X}$. Rather than stack over stations immediately, as is common in back-projection, we first compute graph convolutions directly over this pairwise structure, transforming the latent feature of each (source-station) node with transformed messages from adjacent (source-station) nodes (nodes on the Cartesian product graph $\mathcal{S} \times \mathcal{X}$ being adjacent due to either having nearby stations and fixed sources $\mathcal{E}_{\mathcal{S} \leftarrow \mathcal{S}, \mathcal{X}}$, or nearby sources and a fixed station $\mathcal{E}_{\mathcal{X} \leftarrow \mathcal{X}, \mathcal{S}}$). Graph convolutions on this graph have the key potential of allowing locally coherent information between both subsets of nearby stations, and subsets of nearby sources, to share information simultaneously. Allowing spatial interaction directly on the local pairwise structure in $\mathcal{S} \times \mathcal{X}$ can help suppress side-lobes and the spurious interactions that occur in the input \eqref{eq: input tensor} due to partial alignment of a subset of true arrivals with false source coordinate moveouts. Repeated graph convolutions on this structure allow a gradually expanding receptive field, and incorporate longer-range interactions (i.e., paths) of nodes on the graph.

After this transformation we project the signal on the Cartesian product $\mathcal{S} \times \mathcal{X}$ into the spatial domain, $\mathcal{S} \times \mathcal{X} \longrightarrow \mathcal{X}$ by collapsing (summing) over the station axis. This procedure resembles standard back-projection, however we implement it as a learned bipartite message passing layer (from the original Cartesian product graph to one of its factors); this allows the GNN to first learn how to transform the features into a suitable high-dimensional representation amenable to stacking, while also concatenating the source-station distance offsets within the message-passing layer. This, in turn, allows the operator to learn to augment each source-station feature differently prior to stacking, depending on (vectorial or scalar) distance, which is analogous to the customary practice of distance-dependent weighting in back-projection methods (e.g., \citet{li2020recent}). The GNN then transforms the signal on $\mathcal{X}$ with additional graph convolutions, which include a global summary feature in the message passing convolutions of this module.  Following this, space-time attention prediction layers implement prediction branch $f_2$ from \eqref{eq:functions}. A noteworthy aspect of these predictions is that any set of queried space-time coordinates $\{(\vect{x}, t)\}_q$ can be obtained in a single batched forward call, which all use a trained masked graph convolution attention layer to `interpolate' their predictions from the same latent state already preserved on nearby nodes of the fixed spatial graph, $\mathcal{X}$, during branch $f_1$ of the forward call.

An `inverse' bipartite mapping layer then maps from $\mathcal{X} \longrightarrow \mathcal{S} \times \mathcal{X}$, and we repeat graph convolutions on $\mathcal{S} \times \mathcal{X}$, embed a signal for each arrival in $\mathcal{D}$ by reference to nearby nodes in $\mathcal{S} \times \mathcal{X}$ (i.e., for a given arrival, the K relevent nodes in $\mathcal{S} \times \mathcal{X}$ are the K nearest sources that predict arrivals nearby time $\tau_i \in \mathcal{D}_i$ on station $\vect{s}_i$, or all nodes that predict arrivals within $\epsilon$-distance).  Finally we  predict source-arrival association likelihood using attention with respect to subsets (of the same station, and a fixed source) of the embeddings of the picks in $\mathcal{D}$. This yields prediction branch $f_3$ from \eqref{eq:functions}. These last several modules are complex, however they link the source-arrival predictions to both the earlier space-time source likelihood predictions (via being downstream of the latent signal on $\mathcal{X}$), and by transforming through $\mathcal{S} \times \mathcal{X}$, this causes the predictions for each arrival to be smooth over the station graph. Applying the last attention layer over arrival (feature embeddings) of a common station allows the GNN to focus on selecting only one P and S arrival for each station (and a fixed source), even in cases where there are multiple nearby arrivals (and hence may have similar embeddings). These design choices are made to give the GNN a structure that eases the task of smooth data transformation of picks over heterogeneous station graphs to produce both an aggregate consensus of the source space-time activation likelihoods, and coherent association predictions of each arrival with respect to the global and local network coverage.

\subsection{Labels}

The labels for the source localization prediction and source-arrival association prediction, are given as follows. We choose a relatively wide spatial and temporal bandwidth, $\sigma_{\vect{x}}$, $\sigma_{t}$ (e.g., 30 km, 8 s), and for a set of active sources, $(\vect{x},t) \in \mathcal{Q}$ produced through synthetic simulations, any continuous source coordinate $[\vect{x}_q, t_q]$, is given a label

\begin{equation} \label{eq: label}
   \text{Label}([\vect{x}_q, t_q]) = \max_{(\vect{x}, t) \in \mathcal{Q}} \exp\Bigl(-\frac{(t_q - t)^2}{2\sigma_t^2}\Bigr) \exp\Bigl(-\frac{||\vect{x}_q - \vect{x}||^2}{2\sigma_{\vect{x}}^2}\Bigr)
\end{equation}

\noindent which is just a simple kernel approach to represent each source as a continuous normalized space-time Gaussian distribution. Note that for nearby sources, \eqref{eq: label} uses max-pool to set the label only in reference to the nearest source. This approach provides an explicit continuous label space, either for the fixed coordinates at $\mathcal{X}$, or for arbitrary prediction queries $[\vect{x}_q, t_q]$ throughout continuous $\Omega$. 

We define the labels for the source-arrival associations in a similar manner. Specifically,

\begin{equation}
   \text{Label}([\vect{x}_q, t_q], \tau) = \max_{(\tau.\vect{x}, \tau.t) \in \mathcal{Q}} \exp\Bigl(-\frac{(t_q - \tau.t)^2}{2\sigma_t^2}\Bigr) \exp\Bigl(-\frac{||\vect{x}_q - \tau.\vect{x}||^2}{2\sigma_{\vect{x}}^2}\Bigr)
\end{equation}

\noindent where we let $\tau.[\vect{x}, t]$ denote the true source $[\vect{x},t]$ that generates the arrival $\tau$, which is known, since the catalogs are produced synthetically. We choose to set the temporal bandwidth more narrow for the association label (e.g., $\sigma_t = 5$ s), than for the source temporal bandwidth ($\sigma_t = 8$ s). Future efforts could optimize the choice of kernel sizes by tuning desired validation metrics with respect to this choice.

\subsection{Synthetic Training Data}

Given the defined GNN, we train it in a supervised fashion, using synthetic training data as input, for which the target sources and labels are known exactly. This is similar to the approach taken in the RNN-based PhaseLink \citep{ross2019phaselink} association method. While nothing precludes training our GNN on real data, it is easier to generate synthetic pick catalogs covering a wide diversity of station configurations, source distributions, and pick sets, than it is to assemble a real world dataset of similar quality and diversity. To develop the synthetic catalogs we follow a heuristic scheme that presents the GNN with realistic, complex, and noisy training datasets. Each realization of a day-long catalog is obtained by the following procedure:

\begin{enumerate}
    \item Sample a network realization $\mathcal{S} \sim \text{Gen}(\mathcal{S})$. In our case, (with 50$\%$ probability) we either select a random subset of [100 - 374] stations of the NC seismic network with EHZ channels (including 374 total possible stations between 2000 - 2022, between [35.8, 39.8]$^{\circ}$N and [-123.8, -120.3]$^{\circ}$E); or we select the true subset of stations that was available on a single day between 2000 - 2022 from the same region.
    \item Sample number of events, $N \sim$ [500 - 5,000], sample locations uniformly in $\Omega_{\mathcal{X}}$, sample origin times uniformly in [0, 1 Day].
    \item Compute arrival times on all stations, for all sources and phase types; $t_0 + T_k(\vect{s}, \vect{x}))$. Randomly choose a cutoff distance of $d_{thresh} \sim [15, 500]$ km for each event, and delete arrivals of source-station paths $> d_{thresh} + \epsilon$, with $\epsilon \sim \mathcal{N}(0,30  \text{ km})$ sampled iid. per station and per source, and $\mathcal{N}$ a Gaussian distribution.
    \item Corrupt arrival time data: (i). Add Laplacian noise to moveout times, with $\sigma$ linearly proportional to travel time (e.g., 3 - 5$\%$ travel time), (ii). Randomly delete arrivals from $r$ fraction of stations (e.g., $r = 0.2 - 0.8$) from $\mathcal{S}$, for each event.
    \item Check which sources still produce $> N_{min}$ arrivals on $N_{min}$ unique stations (e.g., $N_{min}$ = 4). Mark these sources as `active', and place in set $\mathcal{Q}$.
    \item Add false arrivals to all stations, randomly placing $n_i \in$ [1,500 - 15,000] false picks for all stations. Also, for all real picks, with $3.5\%$ probability, add a random false `coda' arrival, within [0, 25 s] after the pick time.
    \item Choose between [0,100] times between [0,1 Day] to apply false `spike' picks, to between [1,$N_{max}$] stations, within a $\sim$0.25 s window across all of the network simultaneously. This simulates the real occurrence of network-wide spikes that occur across the NC network on some occasions, due to telemetry issues.
    \item Sample inputs following \eqref{eq: input tensor} for a random time step during the day, $t_0$, create source space-time labels for (i). the fixed grid $\mathcal{X}$, (ii). a random 5000 sample batch  $[\vect{x}_q,t_q]$ of queries in continuous $\Omega$, and for (iii). a random 100 sample batch $[\vect{x}_q,t_q]'$ of queries in continuous $\Omega$, create source-arrival labels (for each phase type $k$).
\end{enumerate}

This procedure produces complex, realistic pick data sets and provides data to train the GNN. Example synthetic pick data is shown in Fig. 1a,b, along with the sources, and theoretical moveout curves of the sources. We note that the complexity of the data is high: some events are observed widely across the network, others on very small numbers of stations. There are plentiful false picks, missing picks, the number of events present in a time-window is ambiguous, and arrival times are not perfectly aligned with theoretical moveout curves. At this stage we do not introduce more realistic correlated sources of noise, such as correlated travel time residuals (between nearby stations along similar paths from the source), or spatial correlations between the false pick and miss rates of stations that could arise from spatially heterogeneous sources of noise such as proximity to the ocean or cities. 

The random selection of $d_{thresh}$ for the moveout cutoff for each event is chosen to show the GNN a large number of both `large moveout' events that are large enough to be recorded over most of the network, and `small-moveout' events that are so small that they are recorded on only a fraction of the network, without biasing it to simply expect a preponderance of small earthquakes, as will be the case with real data. Similarly, the spatial distribution of synthetic seismicity is uniform over the region, and are not biased to focus seismicity on the faults, as would be the case with real data. The theoretical travel times we use in all steps are obtained from the regional USGS 3D P and S wave velocity model \citep{hirakawa2022evaluation}, using the Fast Marching Method \citep{sethian1996fast} in a local Cartesian projection of the WGS84 elliptical coordinate system.

\subsection{Training Procedure}

To train the neural network, we sample mini-batches of 75 random realizations of the synthetic catalog generated by (steps 1 - 7) compute losses on all four objectives simultaneously:

\begin{equation} \label{eq: losses}
    \mathcal{L} = c_1\mathcal{L}_{\mathcal{X}} + c_2\mathcal{L}_{\vect{x}_q} + c_3\mathcal{L}_{(\vect{x}'_q,\vect{\tau}^P)} + c_4\mathcal{L}_{(\vect{x}'_q,\vect{\tau}^S)}
\end{equation}

\noindent and update using Adam Optimizer \citep{kingma2014adam} and automatic differentiation \citep{paszke2017automatic}, with $c = [0.5, 0.2, 0.15, 0.15]$. Each of the losses use the L2-norm averaged over the batch. The modules are assembled in Pytorch Geometric \citep{fey2019fast}. We train by updating $\sim$20,000 times and computing new realizations of synthetic catalogs with every update step, which exposes the model to a diversity of training data during the training process. Notably, due to step (1) in the synthetic data procedure, every sample of the batch uses a different station realization, and hence there are different K-NN station graphs $\mathcal{G}_\mathcal{S}$ for each sample of the batch. The four objectives are easily trained in parallel, because as the GNN learns to correctly predict the source prediction from the static grid, $\mathcal{L}_{\mathcal{X}}$, it gains insights that help it predict the other objectives $\mathcal{L}_{\vect{x}_q}$, $\mathcal{L}_{(\vect{x}'_q,\vect{\tau}^P)}$, $\mathcal{L}_{(\vect{x}'_q,\vect{\tau}^S)}$ as well, due to the strong coupling between these labels. This allows all of these objectives to be trained simultaneously, and links the training of predicting the spatio-temporal source likelihood and discrete source-arrival association likelihoods.

To ensure stable, repeatable training runs of the model, we included a few additional considerations. It was important to ensure that in each batch there are a sufficient number of `positive label' samples (i.e., inputs that actually relate to true sources and are not just noise, with labels equal to zero). We aim for $15 - 30 \%$ positive labels in each batch, and found that much less than this ratio slows training. The ratio of positive to negative labels will also influence the resulting model's tradeoff in recall and precision rates; however, understanding this tradeoff will require extensive re-training and testing, which we leave to future work. Additionally, to avoid large values in the internal feature activations of the GNN, we scale all real-valued quantities such as Cartesian offset vectors between stations and sources (which can be 1000's of meters in SI units in our application), into a range closer to $[0,1]$, which we accomplish using fixed non-trainable scaling vectors at appropriate places in the model. We also found that by using PReLU instead of ReLU throughout the network notably improved the smoothness of training. It required $\sim$5 days to train the network on a single Tesla T4 GPU, with nearly $\sim$2 day of this time spent creating synthetic data between each batch. 

\subsection{Arrival Picking}

For all applications to real data, we first require applying a picker to determine arrival times of possible seismic phases. For this purpose, in this work we use the previously trained U-NET picker, PhaseNet \citep{zhu2019phasenet}, which was trained on 100 Hz seismic data throughout northern California, and is effective at picking both P and S waves. We apply the model for all available EHZ channels of the Northern California (NC) seismic network available during each day of the study. We use only the EHZ channels for picking since the majority of NC stations have EHZ channels available, while a much smaller fraction (e.g., $\sim$15$\%)$ have all three-components available. We use picking thresholds of 0.1 for both P and S wave pick probabilities. We discard the phase label that PhaseNet produces, and treat all arrivals as unlabelled phase types when input to the GNN. When generating the synthetic seismicity catalogs, we tune the pick rates to be comparable to the rates produced by PhaseNet at this threshold.

\subsection{Continuous Time Application}

To process an individual contiguous day with our GNN, we take the picks produced by PhaseNet from whichever set of NC stations (EHZ channel) are available from the NCEDC on that day, and construct the station graph, $\mathcal{S}$, for available stations. We then compute the input (2) at query time $t_0$, followed by predicting a 10 s long spatio-temporal prediction of the source history centered at this time. Next, we step forward by 2 s, repeat the input and prediction steps, and average overlapping portions of the spatio-temporal outputs, until the entire day is processed. We use a single fixed threshold for all days, with points in the continuous space-time output above the threshold 0.3 marked and clustered to individual sources following the local graph-based space-time marching method (described in the Supplemental). For the unique discrete set of obtained sources, we use branch $f_3$ of the forward map (1) and predict association likelihoods of individual picks (and their phase types) to these sources. We retain picks exceeding association prediction threshold 0.2. To select only one phase type for each source-station pair with the highest predicted association values, we use the constrained ILP association routine \citep{mcbrearty2019earthquake} on the GNN output to obtain the final physically consistent set of source-arrival phase assignments. Finally, using these picks, we re-locate each event with travel time based residual minimization (e.g., \citet{tarantola1982generalized}, described in the Supplemental), which produces locations that are consistent with the original GNN location predictions (Fig. S3). For all applications, we consider sources within the spatial region [36.3, 39.3]$^{\circ}$N, [-123.3, -120.8]$^{\circ}$E, and use all NC stations within the slightly expanded region of [35.8, 39.8]$^{\circ}$N, [-123.8, -120.3]$^{\circ}$E.


\section{Results}

To demonstrate the effectiveness of our method, we show results for both synthetic data and real data. For the synthetic data, we confirm that the model smoothly decreases the loss during training, converges (Fig. S2), and makes accurate predictions of synthetic sources for which the ground truth is known (Fig. 5). For example, in Fig. 5, we observe that over a $\sim$1,200 second synthetic catalog interval (not included in training), comprising 22 events, our GNN accurately predicts the majority of the sources. The spatio-temporal outputs produce smooth space-time Gaussian distributions of highly consistent dimensions (set by our label training parameters, $\sigma_t$, $\sigma_{\vect{x}}$). The outputs have a consistent maximum amplitude scale even for closely overlapping earthquakes, and events of different magnitudes (or events of different number of picks), and show very little sign of `side-lobe' type predictions. These features differentiate the GNN's behavior from standard BP methods where side-lobes are significant (e.g., Figs. S6-S9), and output values scale strongly with event magnitude (Table S1). The missed events in Fig. 5 occur only for events with very few corresponding picks. Finally, we find that the synthetic input pick data, which includes a high rate of false picks, does not cause a high rate of false positives for the GNN. Nearly every predicted high-amplitude space-time Gaussian corresponds to a known source.

Fig. 6 shows examples of space-time source predictions, and association predictions for two real earthquakes, using PhaseNet-produced input picks with unlabelled phase types. We note that the obtained source locations are very close (e.g., $<$0.1$^{\circ}$ degree offset) to the reported USGS locations, and there is no evidence of side-lobes in the spatial-temporal predictions. The temporal and spatial source predictions are smooth and closely approximate the Gaussians of the target widths chosen in training. In Fig. 7, we show several example detections for events located at Geysers geothermal field, but of different magnitudes. We observe an increasing aperture of associated phases with increasing magnitude, and very little evidence of spurious associations occurring on distant stations. Additional examples of association results as a function of magnitude and spatial position are given in Figs. S14-S16.

\subsection{Detecting known events}

To assess the efficacy of our model on applications to real data more systematically, we applied the GNN to the time intervals of every earthquake M$>$1 reported by the USGS in our study area, over 500 random days between 2000 - 2022. This allows us to test the model over a representative range of network-coverage, and source-distribution conditions. We processed each day following the approach described in `Continuous Time Applications', and matched each of the USGS earthquakes with the nearest event in our set of recovered sources, for events within a temporal window of $t = 10$ $s$ and spatial window of $0.75\degree$. We emphasize that in this application, we ensure the GNN is not trained or applied with any prior knowledge of these source locations. The predictions are made over a continuous time interval spanning the known source time much longer than the width of the temporal label kernel (e.g., 120 s), and the GNN is free to predict no source, one source, or multiple sources throughout the entire spatial-temporal interval, with no bias to the time or location of the known sources.

Through this procedure, we recover 96.4$\%$ (8501/8819) of the target events. The spatial residuals of matched events are approximately Gaussian (Fig. 8c), with means near zero, and standard deviations of $[0.039\degree \text{ }N, 0.040\degree \text{ }W,  8,390 \text{ m depth}]$. The origin time residuals of $\sim$0.96 $\pm$1.09 s are also approximately Gaussian. The predicted locations from the GNN are consistent with the corresponding locations obtained from standard travel-time-based locations (global unweighted least squares minimization), using the picks (and their phase type assignments) associated by the GNN (Fig. S3). The distribution of recovered events closely follows the known faults in the region, and includes a large proportion of sources at Geysers geothermal field (Fig. 8). It is also notable that the high rate of recovered events, $\sim$96$\%$, is consistent across the 2000-2022 time interval (Fig. S4), showing the methods ability to adapt to changing network geometry, which on a given day included from 150 to 350 stations, and variable station density (e.g., Figs. S11-S13). The magnitudes of recovered events computed using the amplitudes of the GNN associated picks in a locally calibrated magnitude scale (e.g. \citet{hutton1987ml}, described in Supplemental), closely match the ground truth (Fig. 9a,b). The GNN shows an ability to detect both the small and large magnitude events, across the full spatial region, while obtaining association assignments that are consistent with the size of the event and the local network geometry (e.g., Fig. 7, and Figs. S11-S13). As an additional form of indirect validation of the utility of the associated pick data, we apply double-difference earthquake location \citep{waldhauser2000double} directly to the associated pick times returned by the GNN, and find a notable improvement in the expression of seismically active faults in the region (Fig. 10). Additionally, several local quarries are highlighted by clusters of seismicity (Fig. S5).

\subsection{Detecting new events}

In the next test we demonstrate the GNN's ability to detect new earthquakes by applying it to a continuous 100-day interval (October 1, 2017 - January 9, 2018), spanning the occurrence of a $M$4.6 event (2017-11-13, 19:31:29 UTC) on the creeping section of the San Andreas Fault. By running the detections using the same parameters as in the 500 day application, we obtain a catalog (Fig. 11) that has increased rates of microseismicity compared with that reported in the USGS catalog. We detect approximately $\sim$4.2 times more events, with the majority of new events occurring outside of Geysers geothermal field (Fig. 9c,d). This observation is expected, however, because the USGS makes use of the high-density, local Berkeley-Geysers (BG) seismic network, which includes $>$40 additional seismic stations (including borehole stations), to develop their catalog, and these are not included in our analysis. For this reason, the USGS catalog includes many events M$<$1 in this area, and our increased event rates occur at source locations predominately outside Geysers geothermal field (Fig. 9d).

Spatially, we find that the density of new events conforms to the density of events in past catalogs (Fig. 11a,b). The highest density of new events is clearly located at the Geysers geothermal field, followed by activity on most of the major regional faults - including those that are known to creep - and especially in the aftershock region of the November 13, 2017, M4.6 on the creeping section of the San Andreas Fault. Temporally, we observe that rates in the southern portion of our region surrounding the M4.6 epicenter significantly increase following the event, and then decay to background rates following an Omori-decay style curve (Fig. 11c). The estimated magnitudes of new events are small (Fig. 11d), placing them at or below the level of completeness of the standard catalog. Through visual inspection and event timings, we also confirm there is little indication of split events (i.e., falsely duplicated events due to mis-association of phases). Because there is no ground truth reference, we do not attempt to estimate false positive and false negative rates for our new events. However, an example of waveforms detected and associated by our method near the aftershock zone of the M4.6 event (2017-11-13, 19:31:29 UTC) are given in Fig. 12, where we observe expected behavior such as similar duration and style waveforms for fixed source-receiver paths. Additionally, the Gutenberg-Richter (GR) curve of the obtained catalog (Fig. 9d) follows the expected slope at low magnitudes (M$<$1.5), and agrees with the USGS catalog for M$>$1.5, indicating that nearly all new events are below the magnitude of completness of the USGS catalog, and follow a realistic magnitude distribution.

\section{Discussion}

\subsection{Potential of Method}

We have shown the potential of a GNN-based approach for solving the association problem. This is the consistent with recent findings of successful performance of GNN's in earthquake characterization \citep{van2020automated, mcbrearty2022earthquake, zhang2022spatio}, and phase picking \citep{feng2022edgephase}. The method we propose is applicable to conditions that feature highly variable noise levels, uneven and time-variable station coverage, and overlapping sources of different magnitudes. We solve both spatio-temporal source localization and discrete source-arrival assignment using the same forward map \eqref{eq:functions}. The output source predictions are highly regular and smooth in space and time, allowing easy thresholding and clustering to determine a unique number of sources. The association predictions made by the model are conditioned by the source predictions made during the forward call of the network, which enforces consistency across the two outputs (e.g., Fig. 6). The GNN naturally tends to associate picks only from nearby stations for small earthquakes, and includes more distant stations for larger earthquakes (Fig. 7), as it makes use of geometric and network-wide reasoning rather than only travel time alignment to determine association likelihoods. 

A strength of our architecture is that we apply graph convolutions directly to the Cartesian product graph, $\mathcal{S} \times \mathcal{X}$, between stations and sources. This key feature allows `interaction' between both nearby stations, and nearby source coordinates, while not stacking over either axis explicitly. This allows graph convolutions to occur on the object that is most closely related to the observed raw data, while also incorporating the theoretical travel times as an additional component of the input (2). After obtaining a transformed, high-dimensional latent representation on $\mathcal{S} \times \mathcal{X}$, we project into the spatial component (e.g., $\mathcal{S} \times \mathcal{X} \longrightarrow \mathcal{X}$) with an additional learnable mapping, making our source-temporal detection branch of GENIE analogous to a learned back-projection operation. Similarly, the final learnable association module has access to the earlier spatio-temporal source predictions, the misfits between observed and theoretical arrivals, and is influenced by the station geometry encoded in the station graph, $\mathcal{S}$, since it is down-stream of graph convolutions on $\mathcal{S}\times\mathcal{X}$. Hence, the association predictions are able to learn to make use of more nuanced features when determining associations, rather than just travel time alignment of picks along theoretical moveout curves (e.g., \citet{ringdal1989multi}), which sets it apart from non-trainable association algorithms.

We tested the ability of the GNN to recover known seismicity in a `hands off' manner. We trained the network entirely on synthetic data, with no bias of the source distribution to the local tectonic fault structure or true density of sources in the region. Application of GENIE to picks produced by PhaseNet around times of known earthquakes led to recovering $>$96$\%$ of known events M$>$1, with detections that had small spatial and temporal residuals and clearly illuminated the local faults (Fig. 8). The results were robust to time-variable station geometry (Fig. S4), and the associations were of sufficiently quality to allow calibrating a local magnitude scale (Fig. 9), and applying double difference relocation (Fig. 10). The relocated seismicity highlights known local faults and quarries (Fig. S5).

When applying the method to a 100-day interval surrounding a known M4.6 aftershock producing sequence, we observed a background rate of $\sim$4.2$\times$ higher seismicity compared to the USGS catalog outside of the Geysers geothermal field. The new events still follow the expected spatial (Fig. 11a) and magnitude (Fig. 9c,d) distributions, with most events detected near active faults (Fig. 11a,b). The obtained magnitudes yielded a GR-curve with a b-value of $\sim$1.0, that agrees with the USGS catalog above their magnitude of completeness (Fig. 9d). Additionally, we detected a high spatial and temporal concentration of events from the M4.6 aftershock sequence (Fig. 11). Future effort will be to address the quality control of new detections, and to develop a longer continuous time catalog. In all applications, the results of the GNN outputs will depend on the pick data given it. Different pickers could be trained to pick to lower sensitivities, or to reduce the false positive pick rate; each of which could have large effects on the resulting associations and detections.

\subsection{Directions for Improvement}

A number of improvements in our method are possible, including in: (i). the synthetic data generation, (ii). the adjacency matrix (or edge set) creation scheme, (iii). the input representation, and (iv). the internal sub-module complexity. 

Improving the diversity and realism of the training data will be an important direction in improving the performance of GENIE. Generating data that more realistically captures noise conditions (including correlations in travel time residuals, and false and missing pick rates across stations), would reduce the distributional shift between real and synthetic data, and improve the performance of the model. Additionally, constraining the local travel times more accurately, and making use of amplitude or phase type information (e.g., \citet{zhu2022earthquake}) in training would be a valuable direction of improvement. An effective way to include amplitude information along with the input (2) is to measure the misfit of the observed amplitudes with respect to the theoretical amplitudes predicted from a grid of possible magnitude values (e.g., \citet{mcbrearty2022earthquake}), and additionally, the probability of P or S phase classification can be included as an additional input feature.

A more comprehensive hyperparameter search over the adjacency matrix construction scheme used in our method (e.g., varying $K$, or using other edge construction schemes) would be revealing. Multi-scale GNN's have been proposed \citep{xu2021maf}, and other graph types such as `small world' graphs \citep{newman2000models}, expander graphs \citep{hoory2006expander}, and K-NN variants with improved global connectivity guarantees \citep{berton2018comparison} may prove favorable. It is also possible to learn graph generators \citep{you2018graphrnn}. Properties of graphs useful in their analysis and design include their long-range connectivity properties \citep{newman2000models}, the degree distribution of nodes \citep{britton2006generating}, the propensity, arrangement, and type of motifs on the graph \citep{itzkovitz2005subgraphs}, and the spectral properties of the graph Laplacian \citep{merris1994laplacian}. Improved choices of adjacency matrices will directly give GENIE increased capacity and insight. The many ways it is possible to construct the adjacency matrices used reflects the wide functional design space accessible to GNNs.

Rather than use the simple kernel-based input strategy (2) to map the raw picks onto features over $\mathcal{S}\times\mathcal{X}$, it would be valuable to directly pass each node on $\mathcal{S}\times\mathcal{X}$ all observed picks within some misfit of the theoretical arrival time, directly as a set (3b). With each pick, per-pick feature data such as amplitude, phase type, and azimuths could be included, and an additional GNN sub-module could be added at the input to apply graph convolutions to each of these (per station-source pair) input pick subsets. Finally, it would also be possible to pass full-waveform slices of waveforms from each station at each specified lag time (for all source-station pairs), which would make this model a full-waveform based end-to-end prediction model (e.g., \citet{zhu2022end}). The main limitation to implementing this change is simply that realistic full waveform synthetic data at high frequencies (e.g., $>$ 2 Hz) are not easily available, and real training data would likely have to be used instead.

At each sub-module [1-9] of GENIE, the feature dimensions of each individual FCN, the number of repeated convolutions per module, and the type of edge data (and use of global states) could be varied. Our preference has been to favor small feature dimensions (e.g., 15 - 30) and only modest repeated convolution operations (e.g., up to three per module), resulting in a small number of free parameters for the total model (e.g., $\sim$45,000). Increasing the feature dimensions of individual layers and adding features such as batch-normalization or dropout could improve accuracy \citep{zhou2020effective}. Of particular interest should be the use of global states \citep{battaglia2018relational}, long-range shortcuts of latent features between sub-modules (or repeated layers) of the GNN \citep{xu2018powerful}, and node embeddings that make use of absolute node-position and node-type information \citep{you2019position, you2021identity}. These features could allow efficient, compact, long-range transfer of summary information over the graph. Lastly, in the training paradigm itself, meta-objectives \citep{hospedales2020meta}, such as encouraging the uniform suppression of the spatial residual bias field, and information-theoretic regularization of the forward map \citep{tishby2015deep, wu2020graph} could improve generalization between synthetic and real data and enhance accuracy of the outputs.

\section{Conclusion}

Earthquake phase association is a fundamental problem of observational seismology.  It has recently seen revived interest, driven largely by improvements in phase picking that have produced highly increased availability of picks from small earthquakes. These higher rates often lead to close overlap of nearby sources in space and time, making it difficult to unambiguously determine a source history and assign each pick to a source. Many traditional association methods rely on hand-crafted features and heuristic design choices, limiting their adaptability and performance in complex settings. We proposed a graph-based supervised learning method that can carry out both the source and phase association prediction tasks simultaneously. By making use of graphs throughout the architecture we introduce an effective inductive bias into the model that is compatible with the physics of the problem. This trainable associator has the potential to enhance automatic monitoring work flows and enable detecting small seismic sources from complex input streams of picks or waveforms on heterogeneous, time-varying station networks.

\section{Acknowledgments}

The authors thank William Ellsworth, Weiqiang Zhu, Joshua Dickey, Richard Allen, and Albert Aguilar for helpful discussions. This work was supported by AFRL under contract number FA9453-21-9-0054.

\newpage

\bibliography{bib}
\bibliographystyle{apa}

\newpage

\begin{figure}[!htbp] \label{Fig1}
    \centering
    \includegraphics[width=0.995\textwidth]{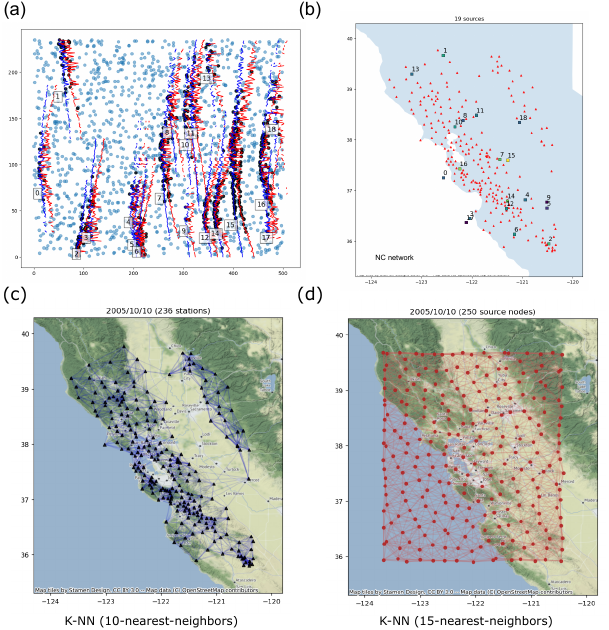}
    \caption{Example synthetic pick data in Northern California (NC), and station and spatial graphs. (a) Raw pick data as observed on NC stations (indices sorted south to north). P and S wave moveout curves are plotted for each event in red and blue respectively, and spatially extend only to the distance of the most distant observing station. (b) Map view of corresponding events in (a), and example set of NC network stations available on a single day. (c,d) Example station and spatial graphs input to the GNN, following the K-nearest-neighbors edge construction scheme.}
\end{figure}

\begin{figure}[!htbp] \label{Fig2}
    \centering
    \includegraphics[width=0.995\textwidth]{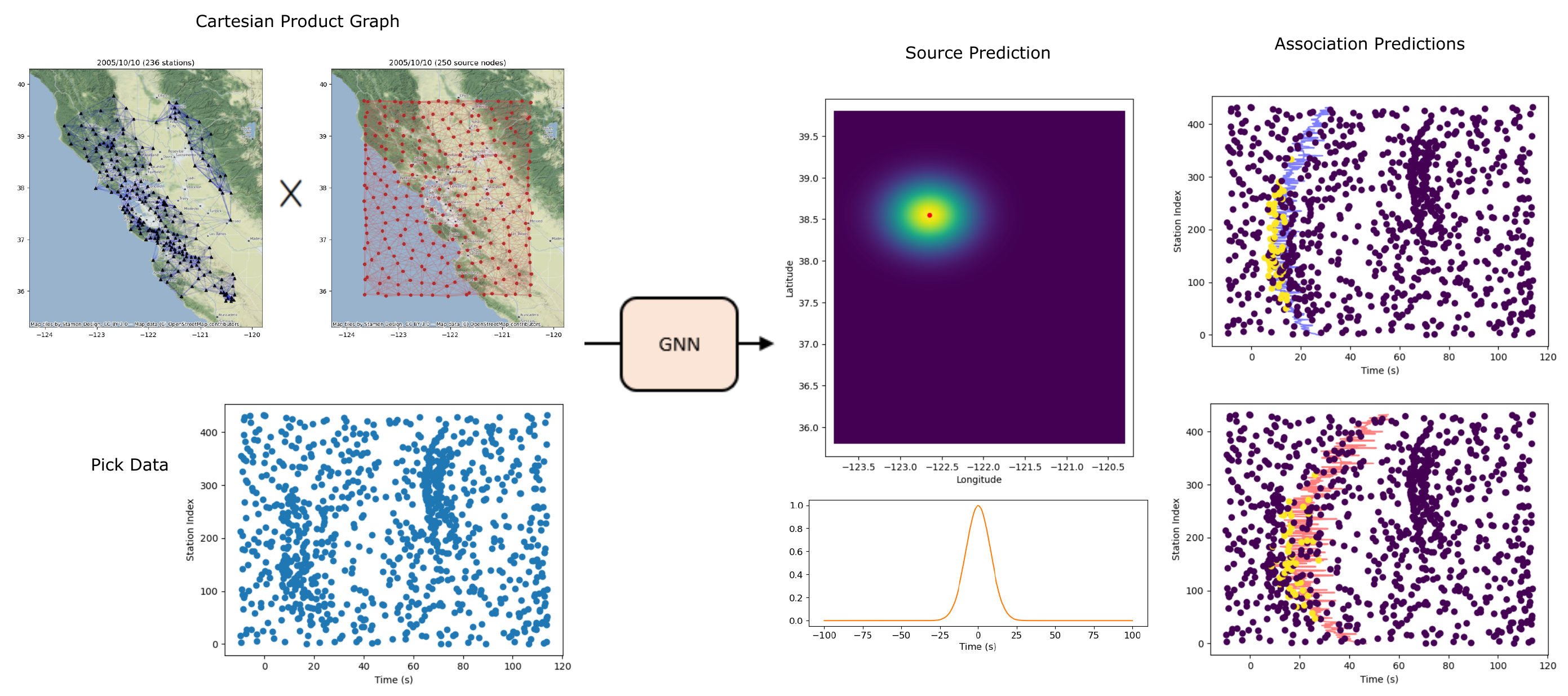} 
    \caption{Schematic of the GNN model's input and output. On the upper left are input station and spatial graphs, and on the lower left a window of input pick times. This input is mapped through the GNN to provide a spatial and temporal prediction of the source likelihood (represented as space-time Gaussians), and individual source-arrival association likelihoods for each pick. Out of the large number of picks within the input pick window, the GNN is trained to identify only the small subset of P (top right) and S (bottom right) picks that are associated to the true source, as marked in red in the spatial heat map.}
\end{figure}

\begin{figure}[!htbp] \label{Fig2}
    \centering
    \includegraphics[width=1.0\textwidth]{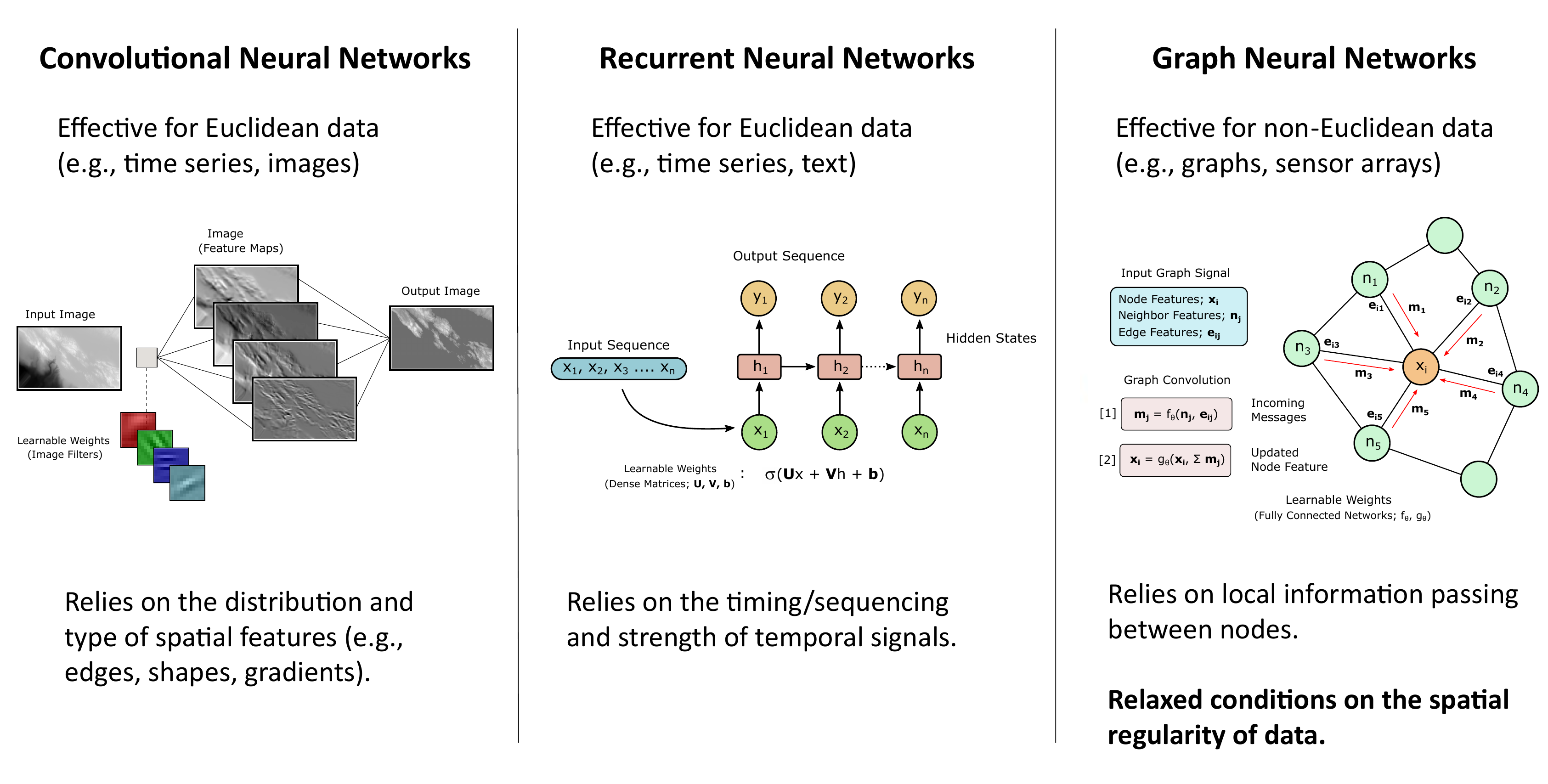} 
    \caption{Schematic comparing the different types of neural networks: Convolutional Neural Networks (CNN's), Recurrent Neural Networks (RNN's), and Graph Neural Networks (GNN's). The mechanisms by which each model transforms data, and the common dataset types they are applied to are shown. }
\end{figure}

\begin{figure}[!htbp] \label{Fig3}
    \centering
    \includegraphics[width=0.95\textwidth]{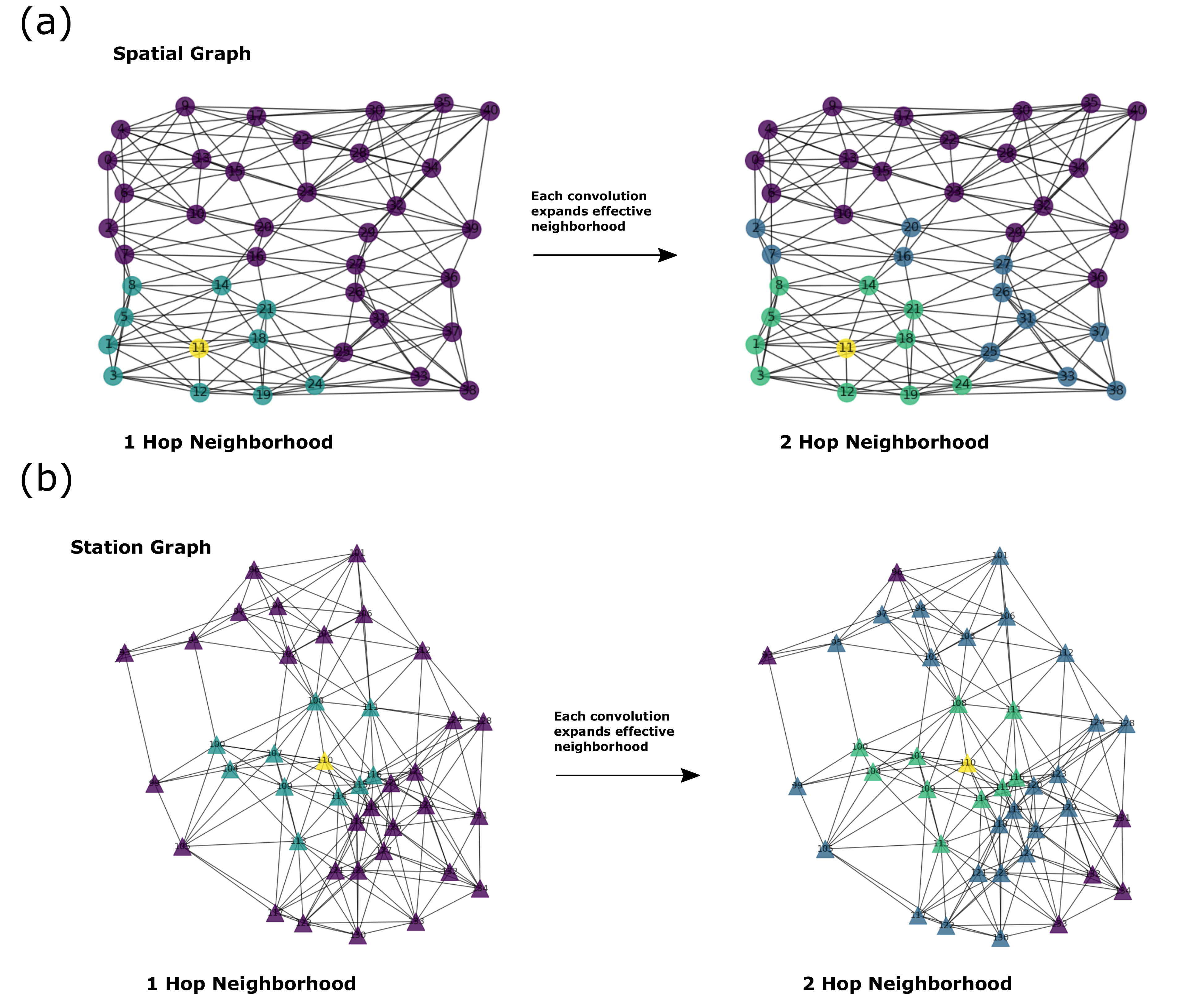}
    \caption{Schematic of graph convolutions, on the spatial (a) and station (b) graphs, for a two-layer GNN application. The central highlighted (yellow) node in each graph is the node receiving messages from neighboring nodes ($j$ $\in$ $\mathcal{N}(i)$). On the left panels, at `1 Hop distance', the partially shaded nodes (green) are direct neighbors of node $i$, and have shared information with node $i$. After a second graph convolution layer is called, the `2 Hop distance' nodes have indirectly shared information to node $i$, since they are linked to the direct neighborhood set, $\mathcal{N}(i)$. Each graph convolution layer results in an expanded 1-Hop receptive field. Further, all nodes have their own receptive fields, and are receiving messages from their own neighbors in parallel (with shared trainable FCN's), while the figure only highlights the receptive field for one representative node. Each node may have a variable cardinality and aperture neighborhood, and hence the graphs may be highly heterogeneous.}
\end{figure}

\begin{figure}[!htbp] \label{Fig4}
    \centering
    \includegraphics[width=0.85\textwidth]{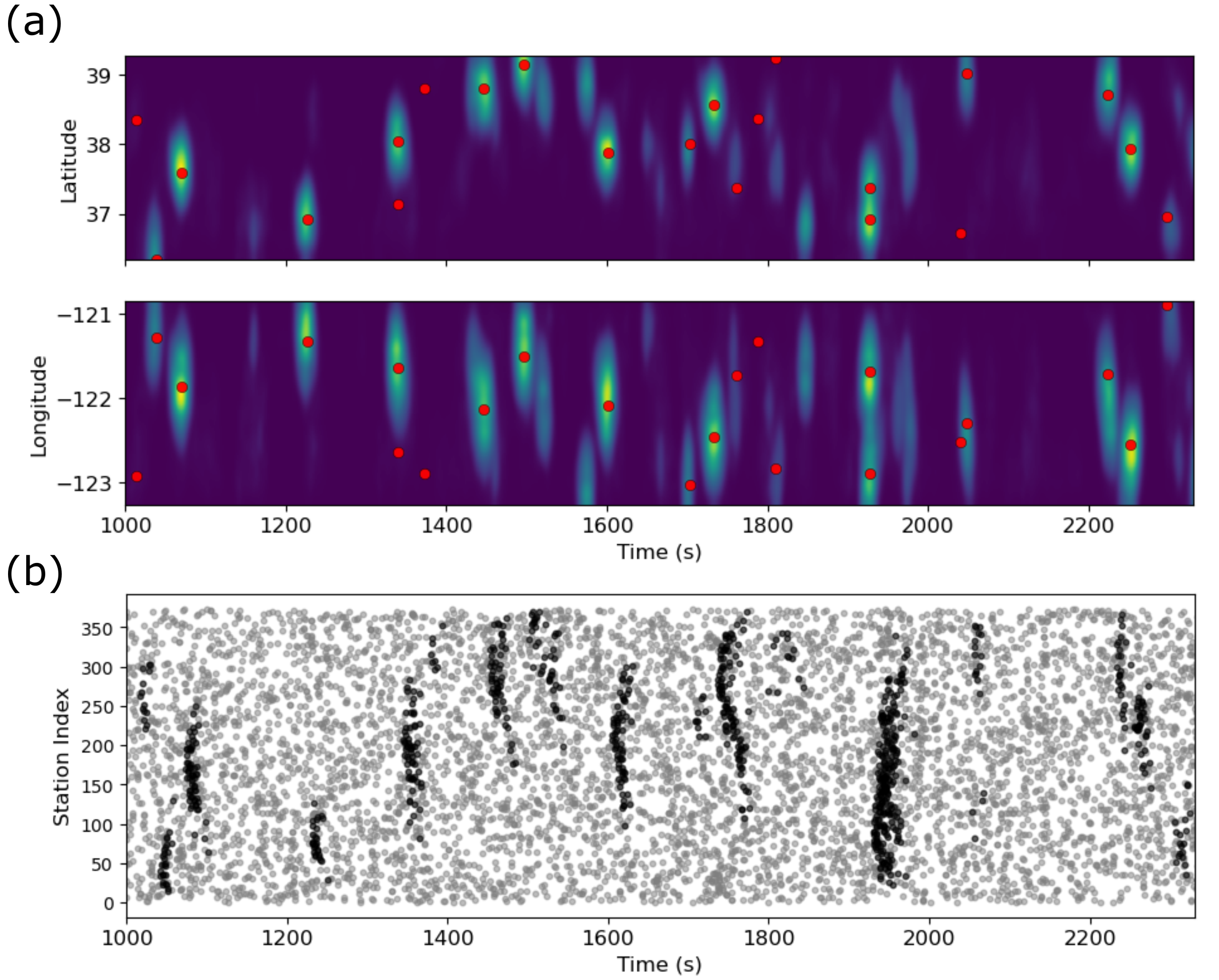}
    \caption{Example synthetic data and spatio-temporal GNN output for a realization not included in training. (a) The spatio-temporal heat maps of the GNN, when marginalizing over the other spatial dimensions. The red dots mark the times and locations of known sources in the synthetic data. (b) The synthetic input pick data. Picks from true sources are black, all other picks in grey are false. At time $\sim$1950 s, two closely overlapping sources are correctly isolated and detected. We note that the visualization in (a) is only for demonstration; in real applications, the output is saved on a continuous 4D grid, and local maxima in the 4D space are extracted from that representation directly.}
\end{figure}

\begin{figure}[!htbp] \label{Fig5}
    \centering
    \includegraphics[width=0.95\textwidth]{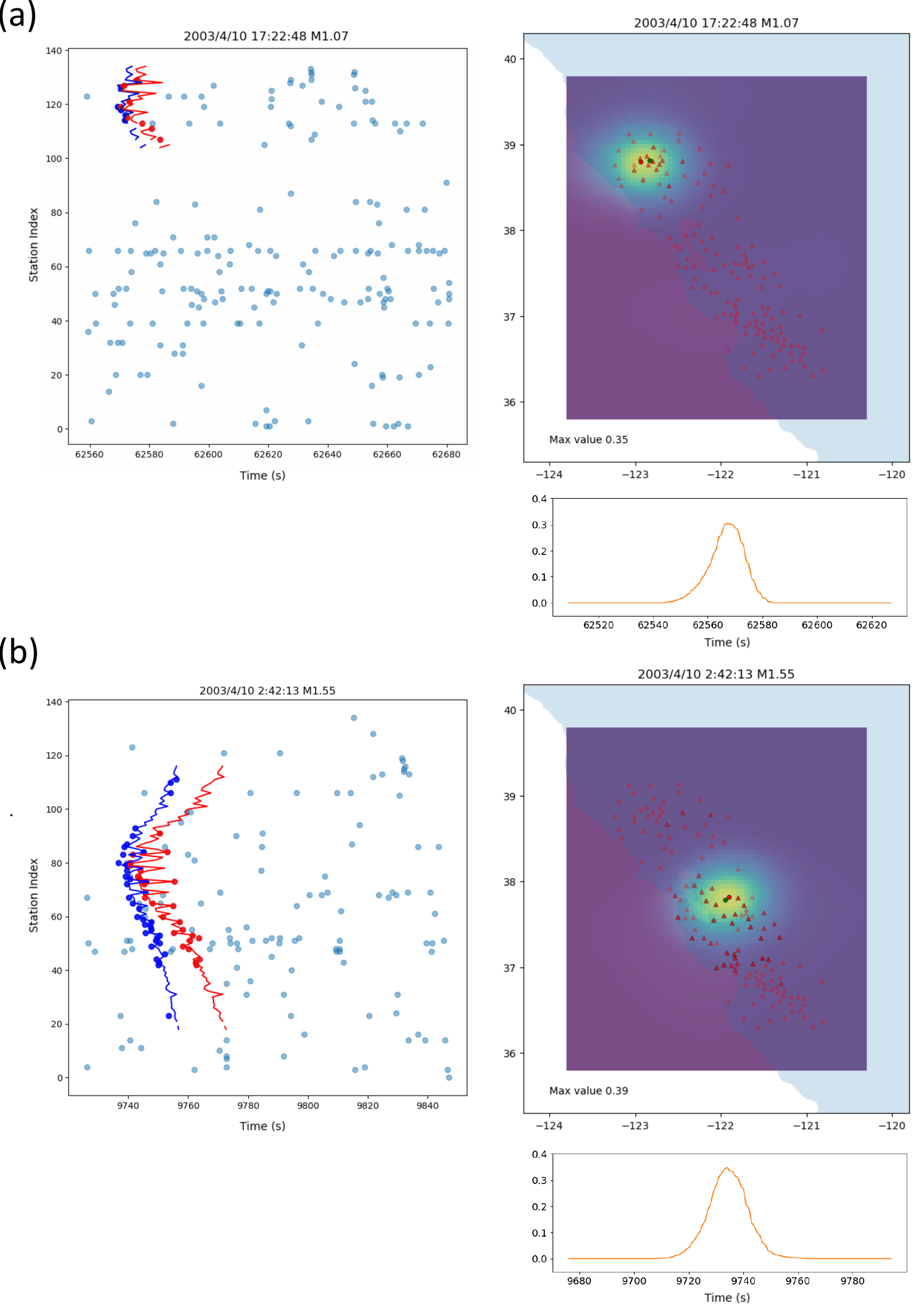} 
    \caption{Example detections of individual real events reported by the USGS. (a) and (b) show the input pick data, spatial heat maps of the predictions, and the temporal likelihood prediction at the coordinate of the maximum point in the spatial domain. On the spatial heat maps, the green dots shows the reported USGS location, and the red dots show our maximum likelihood location. The blue (P waves) and red (S waves) circles in the pick data are the subset of predicted associated picks to the maximum likelihood source coordinate obtained in the spatial maps.}
\end{figure}

\begin{figure}[!htbp] \label{Fig10}
    \centering
    \includegraphics[width=0.95\textwidth]{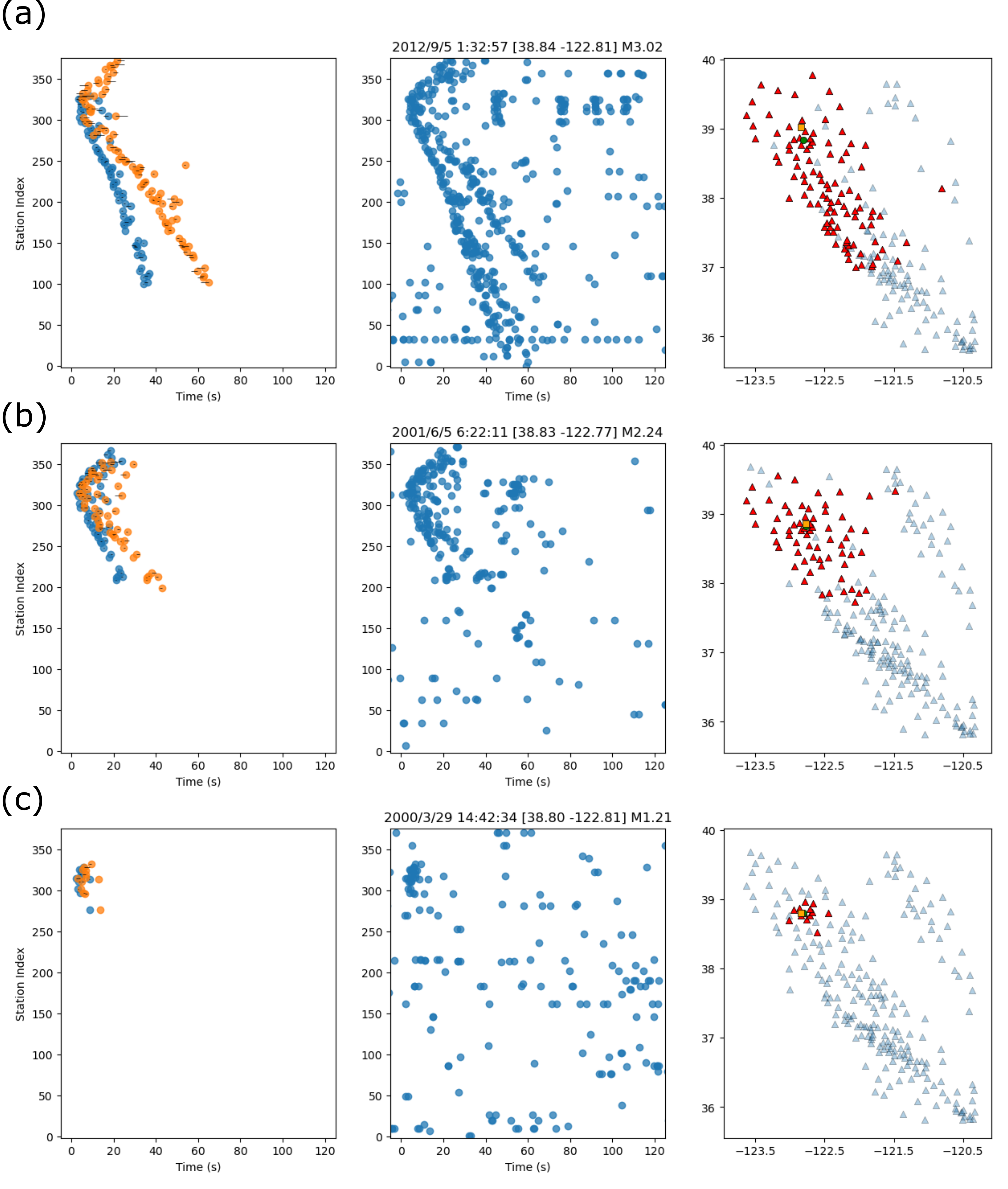} %
    \caption{Example associations across the NC network, for three different events (a)-(c), each corresponding to different magnitude values of approximately $\sim$3.25M, 2.25M, 1.25M, respectively. All events are from similar locations in the Geysers geothermal field. Columns show associated picks (left), initial un-associated input picks (middle), and the obtained active set of stations with an associated pick (right). In right panels, the USGS location (circles) and our location (squares) are marked for each detected source. In left panels, black lines indicate the misfit of the observed arrival from the theoretical arrival time. Misfit lines $<0.35$ s are not visible.}
\end{figure}

\begin{figure}[!htbp] \label{Fig6}
    \centering
    \includegraphics[width=0.95\textwidth]{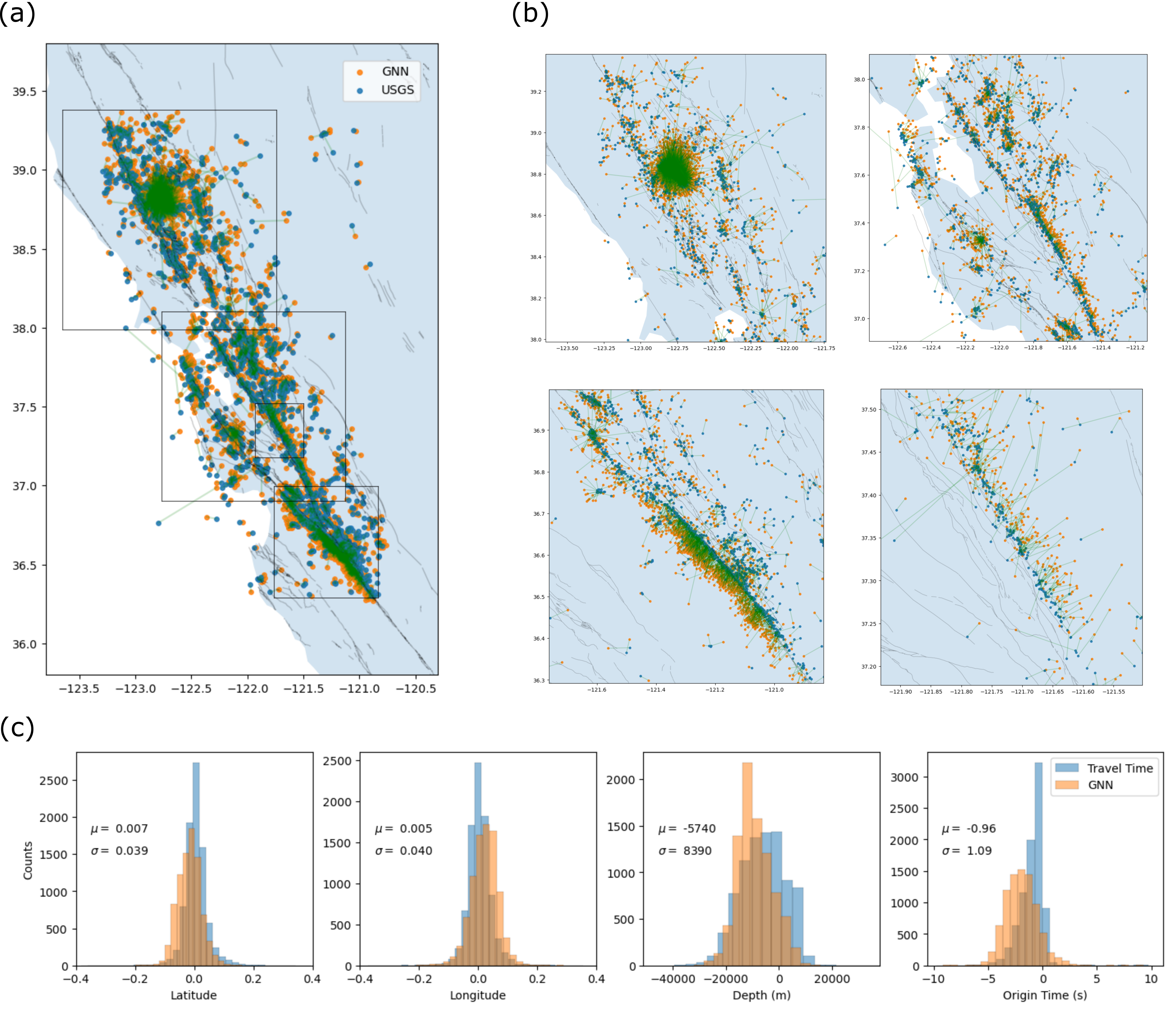}
    \caption{Summary of matched detections from applying the GNN over known event times of M$>$1 events, for 500 random days between 2000 - 2022. Panels (a-b) show the locations of the reference USGS events and ours, where (b) includes the four insets marked in (a). Green lines link matched events. Panel (c) shows the distribution of spatial and temporal residuals of matched events, along with the standard deviations and means.}
\end{figure}

\begin{figure}[!htbp] \label{Fig8}
    \centering
    \includegraphics[width=0.98\textwidth]{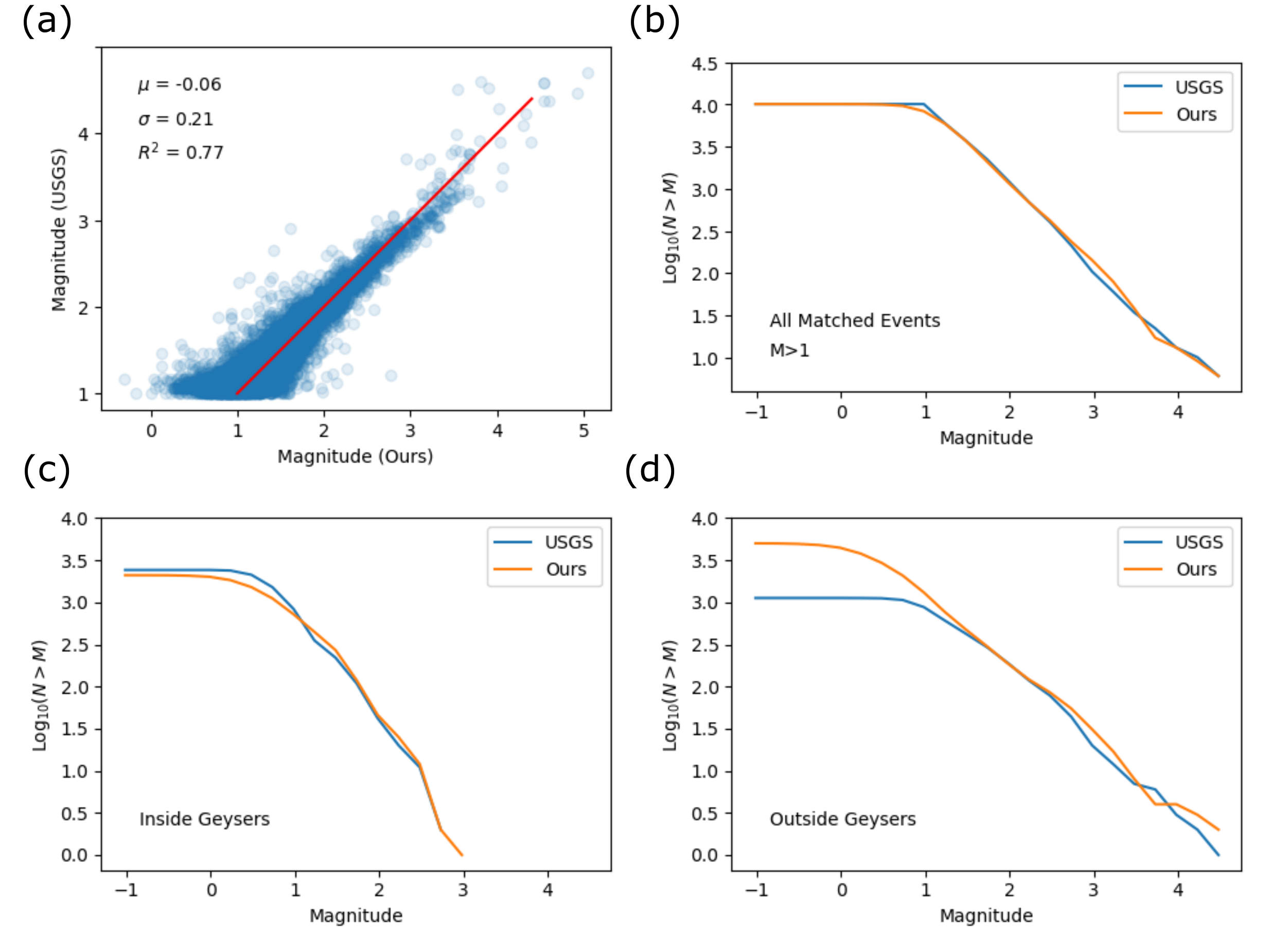}
    \caption{Magnitudes of earthquakes in the USGS catalog and those obtained using our associated picks and calibrated magnitude scale. (a) Cataloged magnitudes compared to predicted magnitudes for matched events during the 500 random day, and 100 continuous day tests. (b) Gutenberg-Richter plots of the USGS catalog and our events for the matched events shown in in (a). (c)-(d) Gutenberg-Richter plots of all the detections obtained during the 100 continuous day test, limited to inside Geysers (region: [38.67, 38.95] N, [-122.94, -122.62] E), and outside Geysers, respectively.}
\end{figure}

\begin{figure}[!htbp] \label{Fig9}
    \centering
    \includegraphics[width=0.95\textwidth]{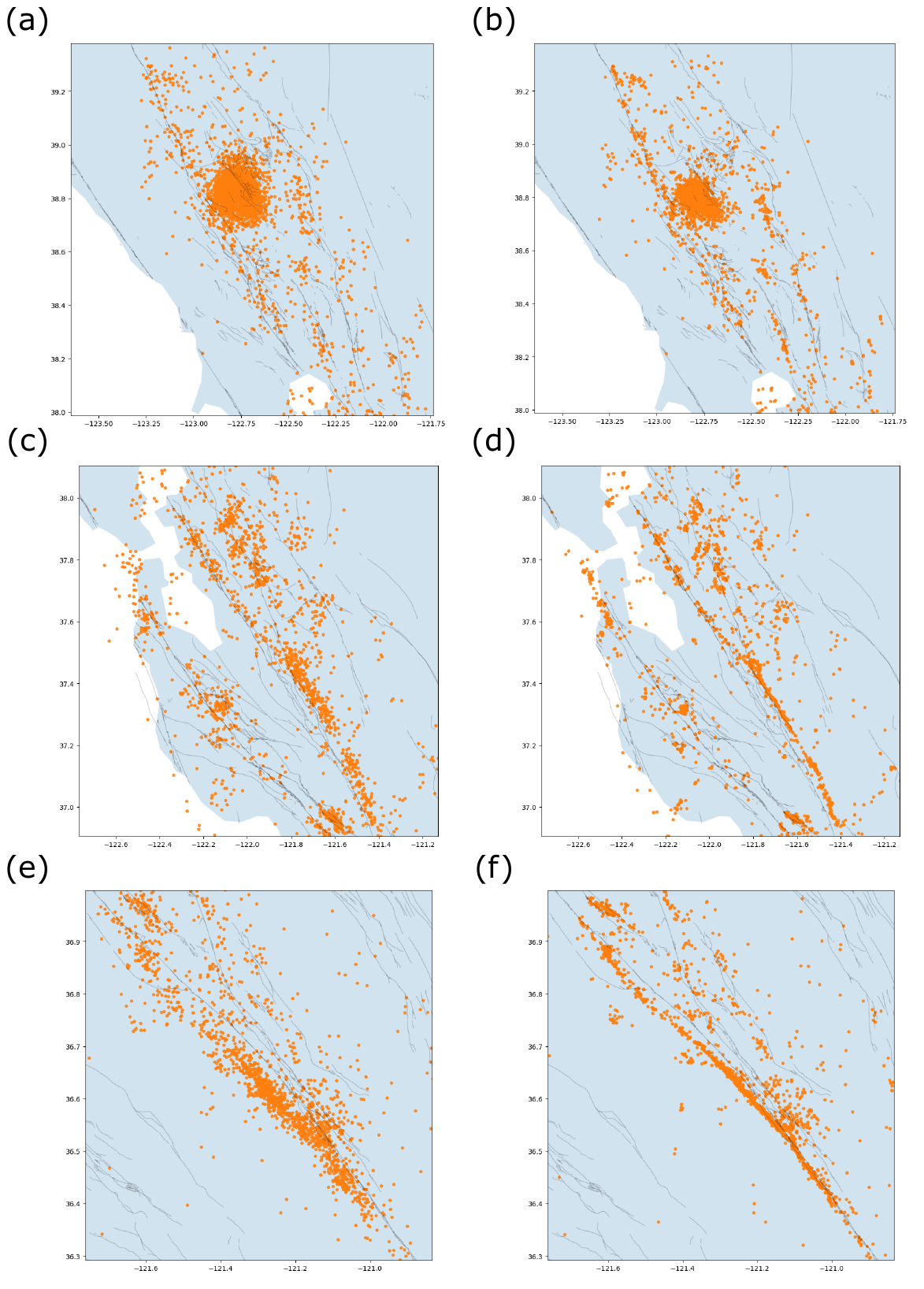}
    \caption{(a,c,e) Travel time based locations of the detected events shown in Fig. 8. These are located using standard (global) least squares minimization of theoretical travel times and the discrete picks associated by the GNN (and also assigned P or S phase type). (b,d,f) The same events as shown in panels (a,c,e), except re-located with double difference relocation \citep{waldhauser2000double} with respect to the original pick times (and no cross-correlation based re-picking).}
\end{figure}

\begin{figure}[!htbp] \label{Fig10}
    \centering
    \includegraphics[width=0.95\textwidth]{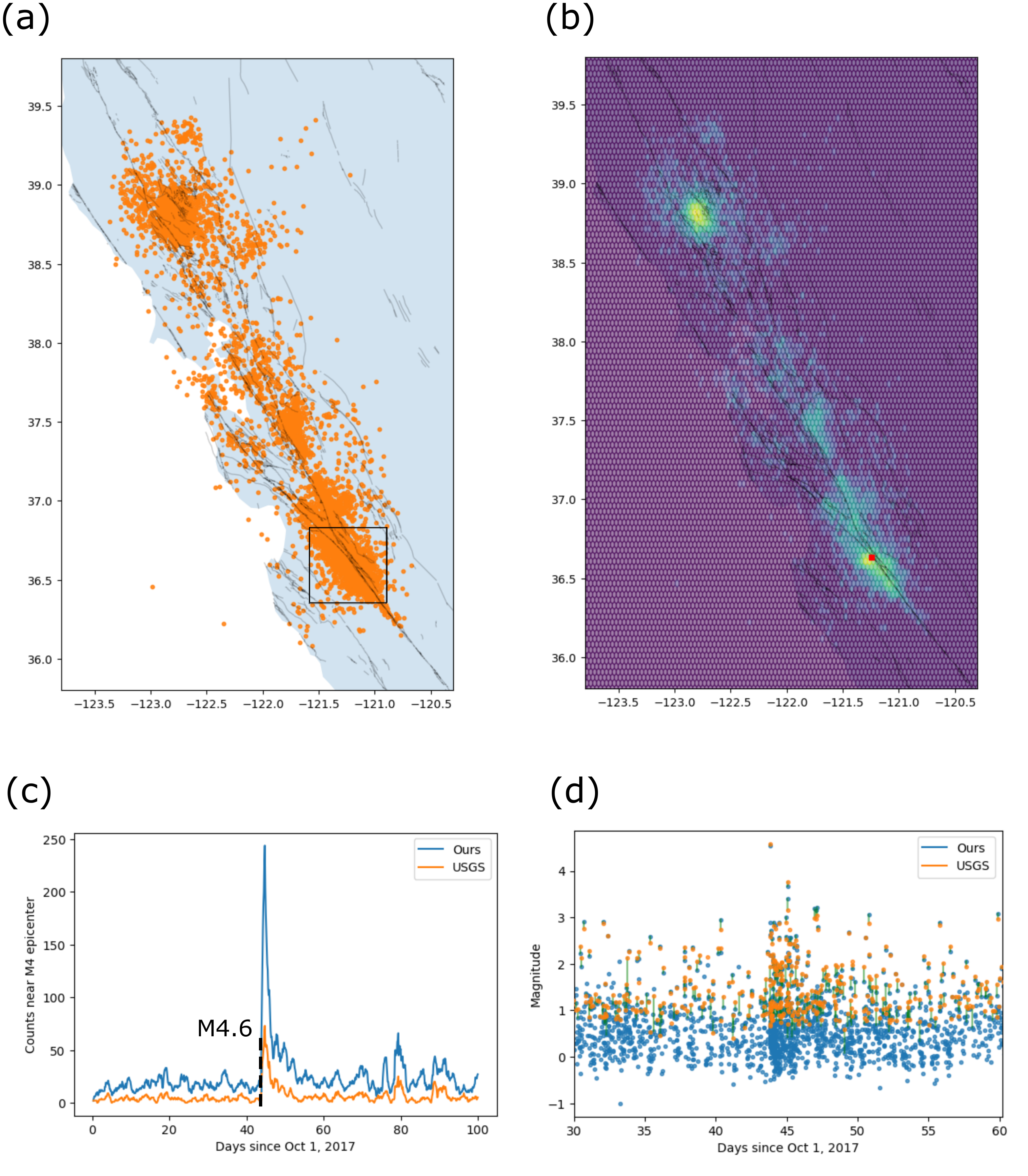}
    \caption{All event detections made during the 100 day interval between October 1, 2017 - January 9, 2018, spanning the time of the aftershock sequence for a M4.6, at 36.63°N 121.24°W, 2017-11-13, 19:31:29 UTC. (a) All detected events, (b) corresponding density plot of events (color scale is in $\log_{10}$ counts), where red square marks the M4.6 epicenter. (c) Counts of detected events per day in the southern spatial window marked by the inset in (a). Time of the M4.6 is marked by the dashed black line. (d). Magnitudes from GENIE vs. USGS catalogs over 30 days in the inset region surrounding the M4.6.}
\end{figure}

\begin{figure}[!htbp] \label{Fig10}
    \centering
    \includegraphics[width=0.99\textwidth]{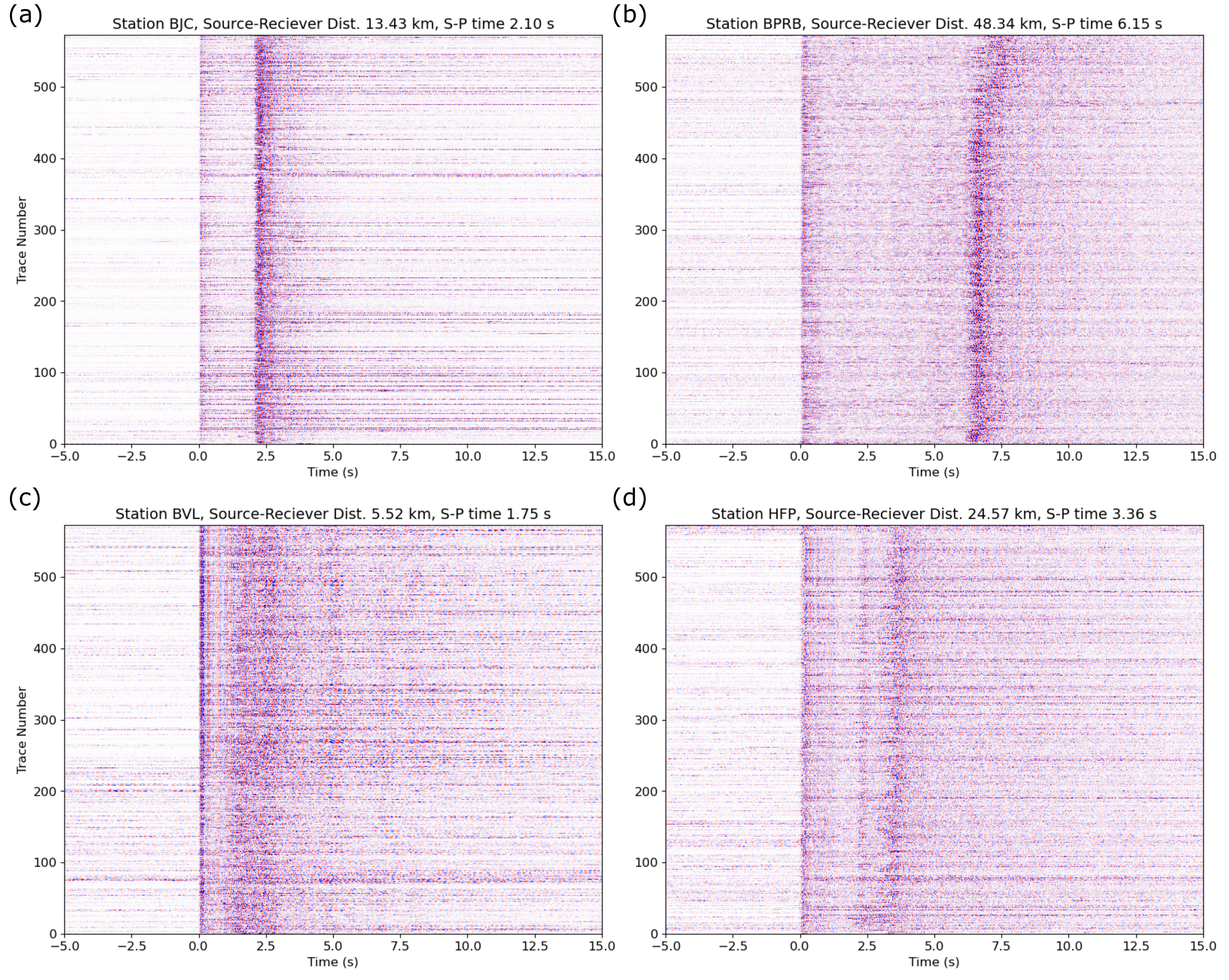}
    \caption{Example waveform sections (a)-(d) aligned on P-wave picks recorded on stations BJC, BPRB, BVL, and HFP of the NC network, and associated to sources within 10 km of the aftershock zone of the M4.6 event shown in Fig. 11. Waveforms are non-instrument corrected, EHZ component velocity traces bandpassed between 2 - 25 Hz, sorted in increasing source-receiver distance, and normalized by the max amplitude values. Approximate source-receiver distances and S-P lag times are listed for each panel.}

\end{figure}

\end{document}